\documentclass[prd,twocolumn,showpacs,nofootinbib]{revtex4}
\usepackage{times}
\usepackage{natbib}
\usepackage{epsfig}
\usepackage{bm}

\newcommand{\bdv}[1]{{\bf #1}}
\newcommand{\mpc}{\rm Mpc}
\newcommand{\kms}{{\rm km\, s}^{-1}}
\newcommand{\mpci}{{\mpc^{-1}}}
\newcommand{\OM}{\Omega_m}
\newcommand{\OB}{\Omega_b}
\newcommand{\up}[1]{{\rm #1}}
\newcommand{\PC}{\varphi}
\newcommand{\kvec}{\bdv{k}}
\newcommand{\beeq}{\begin{equation}}
\newcommand{\eneq}{\end{equation}}
\newcommand{\TT}{T}
\newcommand{\bdi}[1]{\hbox{\boldmath{$#1$}}}
\newcommand{\AVE}[1]{\langle#1\rangle}
\newcommand{\hmpci}{{h\mpc^{-1}}}
\newcommand{\xvec}{\bdv{x}}
\newcommand{\PPM}{P_m}
\newcommand{\qvec}{\bdv{q}}
\newcommand{\GG}{G_u}
\newcommand{\FF}{F_2}
\newcommand{\hmpc}{{h^{-1}\mpc}}
\newcommand{\bear}{\begin{eqnarray}}
\newcommand{\enar}{\end{eqnarray}}
\newcommand{\bng}{\bar n_g}
\newcommand{\kmax}{k_\up{max}}
\newcommand{\kmin}{k_\up{min}}
\newcommand{\gpc}{\rm Gpc}
\newcommand{\hgpc}{{h^{-1}\gpc}}
\newcommand{\pp}{\bdi{p}}
\newcommand{\RA}{\rightarrow}

\begin{document}

\title{Signatures of first stars in galaxy surveys:\\
Multi-tracer analysis of the supersonic relative velocity effect
and the constraints from the BOSS power spectrum measurements}

\author{Jaiyul Yoo$^{1,2}$}
\altaffiliation{jyoo@physik.uzh.ch,~~ jyoo@lbl.gov}
\author{Uro{\v s} Seljak$^{1,2,3,4}$}
\affiliation{$^1$Institute for Theoretical Physics, University of Z\"urich,
CH-8057 Z\"urich, Switzerland}
\affiliation{$^2$Lawrence Berkeley National Laboratory, University of
California, Berkeley, CA 94720, USA}
\affiliation{$^3$Physics Department and Astronomy Department,
University of California, Berkeley, CA 94720, USA}
\affiliation{$^4$Institute for the Early Universe, Ewha Womans University,
120-750 Seoul, South Korea}

\begin{abstract}
We study the effect of the supersonic relative velocity between dark matter and
baryons on large-scale
galaxy clustering and derive the constraint
on the relative velocity bias parameter from the Baryonic Oscillation
Spectroscopic Survey (BOSS) power spectrum measurements. Recent work has
shown that the relative velocity effect may have a dramatic impact 
on the star formation at high redshifts, if first stars
are formed in minihalos around $z\sim 20$,
or if the effect propagates through secondary effects 
to stars formed at later redshifts.
The relative velocity effect has particularly strong signatures in the large 
scale clustering of these sources, including the BAO position. 
Assuming that a small fraction of stars 
in low-redshift massive galaxies retain the memory of the primordial
relative velocity effect, galaxy clustering measurements
can be used to constrain the signatures of the first stars. 
Luminous red galaxies contain some of the oldest stars in the Universe and 
are ideally suited to search for this effect. 
Using the BOSS power spectrum 
measurements from 
the Sloan Data Release~9, in combination with Planck, we derive the upper limit on 
the fraction of the stars sensitive to relative velocity effect $f_\star<3.3\%$ at the 95\% confidence level
in the CMASS galaxy sample.
If additional
galaxy sample not sensitive to the effect is available in a given survey, a
joint multi-tracer analysis can be applied 
to construct a sample-variance cancelling combination, providing a
model-independent way to verify the presence of the relative 
velocity effect in the 
galaxy power spectrum on large scales. 
Such a multi-tracer analysis in future galaxy surveys can greatly improve the
current constraint, achieving a 0.1\% level in $f_\star$. 
\end{abstract}

\pacs{98.80.-k,98.65.-r,98.80.Jk,98.62.Py}

\maketitle

\section{Introduction}
\label{sec:intro}

After the cosmic recombination at $z=1090$, the cosmic background radiation
decouples from the baryon-photon plasma and fades away as the Universe expands
and enters the cosmic dark ages \cite{PLANCK13}.
Until the emergence of the first stars, no light but 21cm radiation
from neutral hydrogens is emitted in the Universe.
Previously, the period of the dark ages to the formation of the first stars
was thought to be well described by simple physics that involves
the linear evolution of matter and baryons
with few chemical reactions and their interaction with photons 
\cite{BALO01,FUOHBR06}.
However, it was shown in \citet{TSHI10} that a typical velocity difference
between baryons and dark matter after the cosmic recombination 
corresponds to a supersonic flow with Mach number 
$\mathcal{M}\sim 2$. During the period $z=15\sim100$, most volume in the 
Universe is filled with supersonic flows and shocks, and the volume with
$\mathcal{M}<1$ is less than 10\%. 
In \citet{DAPESE10} 
it was argued that the effect could dramatically change the 
fraction of baryons captured in dark matter halos, having no baryons 
condensing inside halos when the relative velocity is larger than the halo 
virial velocity, and having baryons inside dark matter halos only in regions 
where the relative velocity is zero.
This and subsequent studies 
\cite{TSBAHI11,FIBAET12,MCOL12,NAYOGN12} 
showed that these supersonic
relative velocities suppress the dark matter halo abundance at low mass,
reduce the gas contents in those halos, and boost the minimum cooling mass,
which altogether delays and suppresses the formation of the early baryonic
structure, by an amount proportional to the local relative velocity. 

High-resolution numerical simulations using adaptive-mesh refinement
or moving-mesh hydrodynamics techniques
\cite{MAKOCI11,STBRLO11,GRWHET11} have confirmed that the relative velocity 
effect influences the virialization of the gas in minihalos and delays the 
population~III star formation.  A series of numerical simulations 
\cite{NAYOGN12,NAYOGN13} provide a statistically robust evidence that
the gas content is significantly reduced below the characteristic mass
scale that is large enough for gas to collapse due to gravity overcoming
the pressure gradient. A further numerical study \cite{OLMC12}
reveals a stunning complexity of gas accretion into the characteristic
mass halos and emphasizes consistent treatment of the relative velocity effect
in simulations.

Furthermore, the presence of the supersonic relative velocity effect at
high redshifts makes the spatial distribution of the star formation 
highly inhomogeneous, boosting the fluctuation signal by
modulating the formation of first 
stars on very large, hundred (comoving) Megaparsec, scales. 
This can be detectable in redshifted 21cm observations 
\cite{DAPESE10,VIBAET12,FIBAET13} at $z>10$ (see also \cite{BILO11,MCOL12}).
More importantly, this spatial inhomogeneity
is modulated by the relative velocity effect that is coherent on larger
scales than the matter density, exhibiting a prominent acoustic oscillation
structure in the power spectrum \cite{DAPESE10}. At high redshifts,
the relative velocity contribution to the power spectrum is substantially 
larger than the matter fluctuation contribution, especially on the baryonic
acoustic oscillation (BAO) scales.

However, 
the relative velocity effect
decays with redshift once the halos that are collapsing have 
virial velocities larger than the relative velocity between baryons and 
dark matter. Even though
the direct impact of the relative velocity effect on the low-redshift
massive galaxies is expected to be negligible, stars that have been formed 
at high redshifts are still expected to retain the memory of
the spatial modulation caused by the relative velocity effect. There are further 
indirect ways how the effect can be important for stars that are forming 
at or after reionization, long after the minihalos ceased to be important
\cite{DAPESE10,GRWHET11,YODASE11,FESMDV12}.
Such scenarios include a 
patchy reionization partially driven by minihalos with the relative velocity effect or
inhomogeneous metal enrichment of the first stars
in the intergalactic medium, both of which
subsequently affect the formation of massive galaxies at late times.

These possibilities imply the spatial modulation of galaxy clustering due to
the remaining relative velocity effect. It is shown \cite{YODASE11} that
due to its acoustic oscillation structure at the BAO scale,
the relative velocity effect, if unaccounted for,
can bias the measurements of the BAO peak
position, but if accounted for, its signatures are sufficiently different that 
it does not bias the estimate of
the dark energy equation-of-state or inflate the parameter constraints.
Furthermore, 
the presence of the relative velocity effect in galaxy clustering can
be unambiguously identified in the galaxy bispectrum on large scales, although 
those constraints are not as strong as direct power spectrum constraints.

Here we extend the galaxy clustering analysis  \cite{YODASE11}
of the relative velocity effect and investigate the synergy effect of 
the multi-tracer analysis.
The multi-tracer analysis \cite{SELJA09} is developed to take advantage
of the fact that the same underlying matter distribution is probed by different
tracers and the sampling variance stochasticity can be completely removed by constructing
a particular combination of different tracers. 
This technique has been applied to constraining the growth of structure
from the redshift-space distortion \cite{MCSE09,BECA11,CABESH11,HASEDE12}
and measuring the relativistic effect and the primordial non-Gaussianity
\cite{MCDON09,HASEDE11,YOHAET12}. We use the multi-tracer analysis to
improve the constraint on the relative velocity effect and provide a 
model-independent way to verify its presence in the galaxy power spectrum.
Moreover, we apply our formalism of the relative velocity effect in galaxy
clustering to the recent galaxy power spectrum measurements
\cite{ANAUET12} from the 
Baryonic Oscillation Spectroscopic Survey (BOSS; \cite{SCWHEI10})
to derive for the first time the constraint on the fraction of the first
stars contained in the galaxy sample.

The organization of the paper is as follows. In Sec.~\ref{sec:relvel},
we briefly summarize the formalism for computing the relative velocity effect
and contrast the acoustic oscillation structures in the matter and the
relative velocity distributions. In Sec.~\ref{sec:effect}, we present a
comprehensive study of the relative velocity effect in galaxy clustering.
In the presence of the relative velocity effect, 
the galaxy power spectrum and the cross-power spectrum are computed
in Sec.~\ref{ssec:pow} and Sec.~\ref{ssec:cross}, and a particular combination
of two galaxy samples is constructed to eliminate the leading-order
sample variance in Sec.~\ref{ssec:multi}. A full covariance matrix
of the auto- and the cross-power spectra is computed in Sec.~\ref{ssec:cov},
and the shift in the BAO peak position is quantified for the auto- and
the cross-power spectra in Sec.~\ref{ssec:shift}. The relative velocity
effect on the auto- and the cross-correlation functions are studied in
Sec.~\ref{ssec:temp} and Sec.~\ref{ssec:corr}.

In Sec.~\ref{sec:boss}, we analyze the BOSS power spectrum measurements
and derive for the first time the constraint on the relative velocity effect
in the galaxy sample. Looking to the future, we forecast the constraint
on the relative velocity effect, providing a guidance to the multi-tracer
analysis in future galaxy surveys in Sec.~\ref{sec:future}.
Finally, we summarize our results and discuss the implications of our results
on relating the current and the future constraint to the primordial galaxy
formation history in Sec.~\ref{sec:discussion}. For numerical calculations, 
we adopt a flat $\Lambda$CDM universe with the fiducial cosmological 
parameters consistent with the recent analysis of the {\it Planck}
satellite mission \cite{PLANCK13}:
The scalar spectral index $n_s=0.968$ and its running 
$\alpha_s=0$ of the primordial curvature power spectrum with
its normalization $A_s=2.46\times10^{-9}$ at $k_0=0.002~\mpci$,
the matter density $\omega_m=\OM h^2=0.140$, 
the baryon density $\omega_b=\OB h^2=0.0222$, the dark energy density
$\omega_\up{de}=\Omega_\up{de}h^2=0.328$ (the Hubble parameter
$h^2=\omega_m+\omega_\up{de}=0.68^2$).

\begin{figure*}
\centerline{\psfig{file=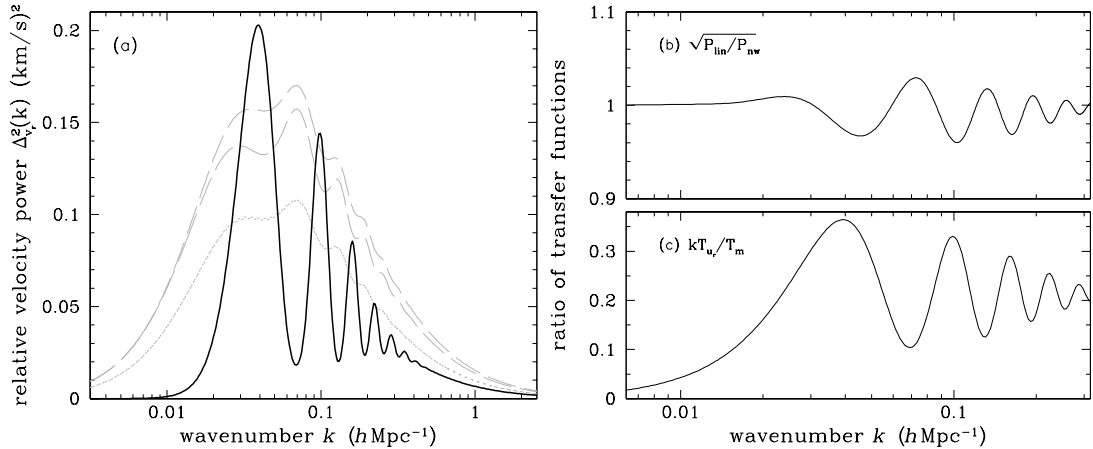, width=3.in, angle=-90}}
\caption{Relative velocity power spectrum and its transfer functions.
{\it Left:}
 Solid curve shows the dimensionless relative velocity power spectrum 
$\Delta_{v_r}^2(k)$ at $z=15$, and the rms relative velocity is 
$\sigma_{v_r}\simeq0.3~\kms$. Velocity power spectra $\Delta_v^2(k)$ and its
scaled power spectra 
$\Delta_\delta^2(k)/k^2$ for baryon (upper) and dark matter
(lower) are shown as gray dotted and dashed curves, respectively
(two gray dotted curves largely overlap with each other).
Their amplitude is arbitrarily reduced in comparison to $\Delta_{v_r}^2(k)$.
While the relative velocity (solid) is sourced by velocities (dotted),
its oscillation is {\it not} in phase with baryon and dark matter velocities.
{\it Right:} Acoustic oscillation structure at $z=0.57$ in the matter transfer 
function~$(b)$ and in the relative velocity transfer function~$(c)$.
The oscillation amplitude ($\sim100\%$)
is significantly larger in the relative velocity
transfer function than in the matter transfer function ($\sim5\%$).
}
\label{fig:transfer}
\end{figure*}

\section{Relative Velocity between Baryons and Dark Matter}
\label{sec:relvel}
We begin by defining our convention for various transfer functions
that are used in our calculations. In linear theory,
perturbations at each wave mode
grow only in amplitude given the initial conditions, such that
its stochastic nature of perturbations
can be separated from the deterministic growth.
The former is characterized by the curvature perturbation~$\PC(\kvec)$
in the comoving gauge, when the wave mode exits the horizon during the 
inflationary period. For the Gaussian initial conditions, the primordial
curvature is fully characterized by its power spectrum
\beeq
\Delta_\PC^2(k)={k^3\over2\pi^2}P_\PC(k)\equiv 
A_s\left({k\over k_0}\right)^{n_s-1}~,
\eneq
where the normalization amplitude is $A_s=2.46\times10^{-9}$ at the pivot scale
$k_0=0.002~\mpci$ and the spectral index is $n_s=0.968$,
consistent with the recent Planck results \cite{PLANCK13}. The deterministic
growth from the initial conditions is captured by the transfer function 
$\TT(k,z)$ for each perturbation variable, e.g., 
\beeq
\delta_m(\kvec,z)=\TT_m(k,z)~\PC(\kvec)~,
\eneq
for the dark matter density at~$z$ (similarly the subscript~$b$ is used
for baryons).
The transfer functions depend solely upon the wave amplitude and redshift.
In practice,
we use the public Boltzmann code {\footnotesize CMBFAST} \cite{SEZA96} 
to compute the transfer functions for dark matter $\TT_m(k,z)$
and baryons $\TT_b(k,z)$.

The velocities of dark matter and baryons are also
computed in a similar manner. Assuming no vorticity $\nabla\times\bdi{v}=0$,
the velocity vector is fully described by its 
scalar perturbation
\beeq
\bdi{v}_m(\kvec,z)=-{i\kvec\over k}~v_m(\kvec,z)~,
\eneq
and the scalar part is again computed by using the transfer function
from {\footnotesize CMBFAST}
\beeq
v_m(\kvec,z)=\TT_{v_m}(k,z)~\PC(\kvec)~,
\eneq
which defines the velocity transfer function for dark matter.
The conservation equation yields that the velocity transfer function 
is related to the matter transfer function as
\beeq
{d\over d\tau}\TT_m(k,z)=-k~\TT_{v_m}(k,z)~,
\eneq
where $\tau$ is the conformal time.
The relative velocity of baryons and dark matter is defined as
$\bdi{v}_r=\bdi{v}_b-\bdi{v}_m$, and  so is the transfer
function $\TT_{v_r}=\TT_{v_b}-\TT_{v_m}$. We often use the dimensionless
relative velocity 
\beeq
u_r(\kvec,z)\equiv{v_r(\kvec,z)\over\sigma_{v_r}}
={\TT_{v_r}(k,z)\over\sigma_{v_r}}~\PC(\kvec)
\equiv\TT_{u_r}(k,z)~\PC(\kvec)~,
\eneq
normalized by the one-dimensional root-mean-square of the relative velocity
\beeq
\sigma_{v_r}^2(z)={1\over3}\AVE{\bdi{v}_r\cdot\bdi{v}_r}
={1\over3}\int d\ln k~~\TT_{v_r}^2(k,z)~\Delta_{\PC}^2(k)~.
\eneq

 At the recombination, the relative velocity between dark matter
and baryons is $\sim30~\kms$, and it strictly falls off as $1+z$ as the
universe expands, since they both respond to gravity in the same way.
Figure~\ref{fig:transfer}$a$ shows the relative velocity power spectrum
at $z=15$. 
We see that it retains 
prominent oscillation structure. 
Gray dotted curves show the velocity power spectra for dark matter and
baryons (almost identical)
at the same redshift but with the amplitude arbitrarily
reduced by $10^5$ to
compare its oscillation structure with the relative velocity power spectrum
(solid). For further comparison, gray dashed curves show the scaled matter
power spectrum $\Delta_m^2(k)/k^2$ (lower) and the scaled baryon power
spectrum $\Delta_b^2(k)/k^2$ (upper). The difference of their time derivatives
is proportional to the relative velocity power spectrum $\Delta_{v_r}^2(k)$
(solid). The oscillation
phase of the relative velocity power spectrum is different from the the
velocities of dark matter and baryons, despite their common origin.

Figure~\ref{fig:transfer}$b$ and~\ref{fig:transfer}$c$ 
further compare the oscillation structure
in the matter transfer function and the relative velocity transfer function
at $z=0.57$. In Fig.~\ref{fig:transfer}$b$, the linear matter power spectrum 
is divided by the no-wiggle power spectrum \cite{EIHU98} to highlight the
oscillation structure without the broad-band shape of the matter power
spectrum.\footnote{The ratio of the matter power spectrum
to the no-wiggle power spectrum is plotted in Fig.~1$b$ of \citet{YODASE11}.
The $x$-axis in their Fig.~1$b$ is in units of $\mpci$, instead of $\hmpci$, 
while its bottom panel has the $x$-axis in units of $\hmpci$, as is
indicated.} The oscillation amplitude is rather small $\sim5\%$ in the
matter transfer function, while the oscillation amplitude is fractionally
order one in the relative velocity transfer function in 
Fig.~\ref{fig:transfer}$c$. The acoustic oscillation in the relative velocity
power spectrum is {\it not} in phase with the matter power spectrum.
Of course this ratio shows the maximal effect at $z=0.57$ and in reality the effect 
will be much smaller since only a small fraction of stars is modulated by the 
relative velocity effect.

\section{Relative Velocity Effect in Galaxy Clustering}
\label{sec:effect}
The relative velocity between dark matter and baryons allows separation
between baryons and dark matter within halos. Especially at high redshift,
when the dark matter halos have shallow gravitational potential,
the relative velocity can be supersonic, advecting baryons out of dark matter
halos and preventing the formation of baryonic structure 
\cite{TSHI10,TSBAHI11}. A series of numerical simulations have verified
the relative velocity effect on the early structure formation
\cite{STBRLO11,GRWHET11,MAKOCI11,NAYOGN12,NAYOGN13,MCOL12,OLMC12}.
The suppression or the formation of the early baryonic structure is therefore
spatially
modulated by the relative velocity effect, and its correlation length is
set by the acoustic oscillation of the relative velocity \cite{DAPESE10}.

The presence of the relative velocity effect can affect the spatial clustering
of the low-redshift massive galaxies, where galaxy clustering measurements
are performed. However, it is more difficult in this case to compute the
amplitude of the relative velocity effect or to predict the existence of
the relic memory of the primordial relative velocity effect. 
Here we present various
ways to identify the relative velocity effect in galaxy clustering
measurements and discuss the impact of the relative velocity effect on
the BAO measurements, if the relative velocity effect persists.

\subsection{Power Spectrum}
\label{ssec:pow}
Following the procedure in \citet{YODASE11}, 
a galaxy population at low redshift is modeled
with two nonlinear galaxy bias parameters $b_1$ and $b_2$ and the
relative velocity bias parameter $b_r$ as
\beeq
\delta_g(\xvec)=b_1~\delta_m(\xvec)+{b_2\over2}~\left[\delta_m^2(\xvec)-
\sigma_m^2\right]+b_r\left[u_r^2(\xvec)-\sigma^2_{u_r}\right]~,
\label{eq:bias}
\eneq
where the relative velocity $u_r=|\bdi{u}_r|$ is computed at the linear
order and the matter density is computed to the second order in perturbations.
The relative velocity bias parameter characterizes our ignorance of the
relic amplitude of the remaining relative velocity effect 
in low-redshift galaxy populations, and the nonlinear galaxy
 bias parameters are
the lowest order coefficients of the local matter density expansion
that relate to
the galaxy number density (the third-order galaxy bias parameters
$b_3$ and subsequent terms are ignored).

\begin{figure*}
\centerline{\psfig{file=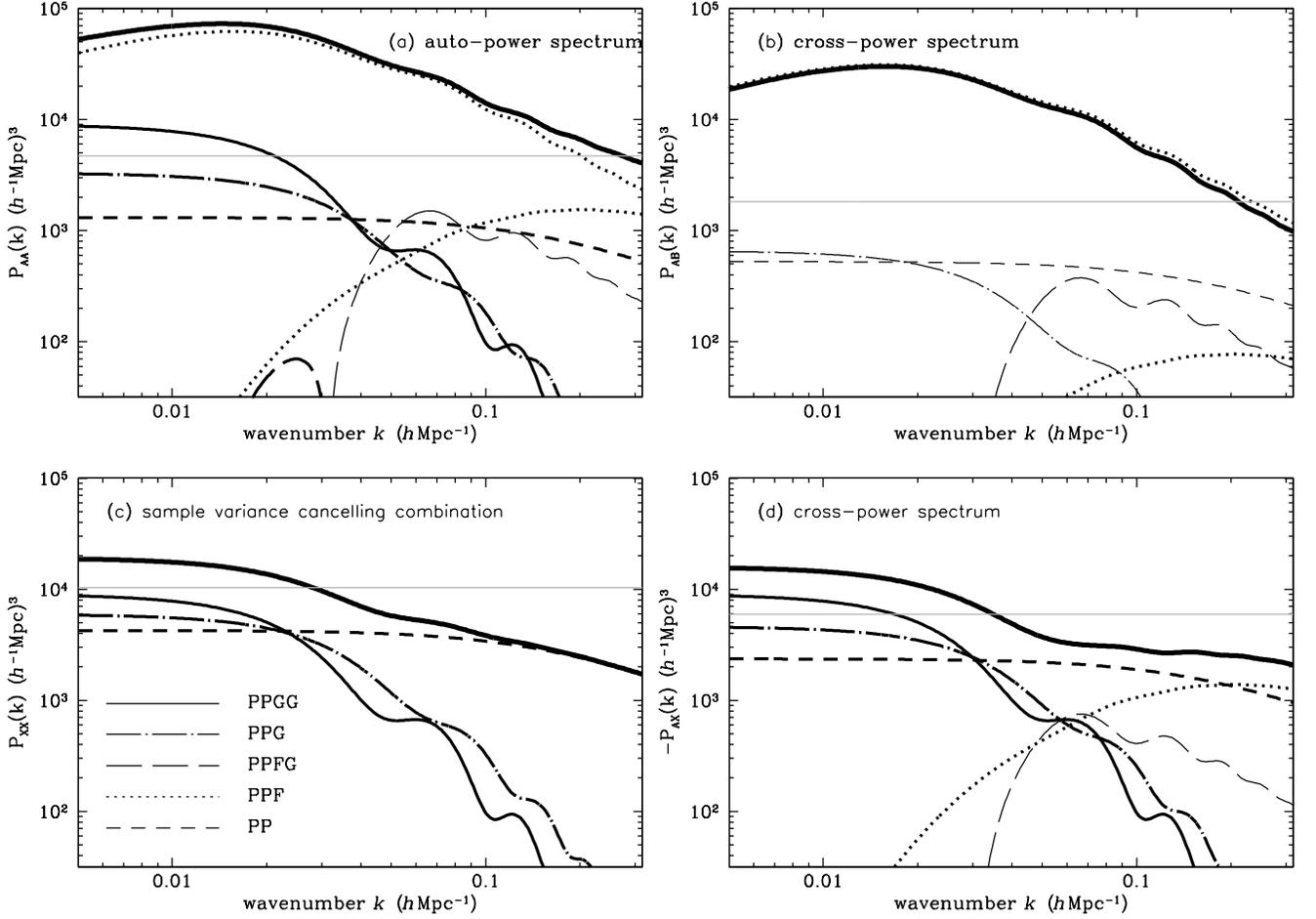, width=5.4in, angle=-90}}
\caption{Auto and cross power spectra of two galaxy samples.
The galaxy sample~$A$ is assumed to have bias parameters
($b_1=2$, $b_2=1$, $b_r=0.04$) and the galaxy sample~$B$ has bias parameters
($b_1=1$, $b_2=-0.4$), but without the relative velocity effect ($b_r=0$). 
Top panels show the auto (left) and the cross
(right) power spectra (thick solid) and their individual contributions (various
curves). Dotted curves that closely follow the thick curves show the nonlinear
matter power spectrum. Two galaxy samples can be properly combined as in
Eq.~(\ref{eq:mulng})
to construct another tracer~$X$, cancelling the sample variance in
Eq.~(\ref{eq:delX}). Bottom panels
show the auto (left) power spectrum of the tracer~$X$ 
and its cross (right) power spectra with the galaxy sample~$A$.
Individual contributions to the total power spectra are shown 
in all panels as
solid, representing the contribution that involves the integral of
$b_r^2\PPM(\qvec)\PPM(|\kvec-\qvec|)\GG^2(\qvec,\kvec-\qvec)$, dot-dashed
for $b_r\PPM\PPM\GG$, long dashed for $b_r\PPM\PPM\FF\GG$, 
dotted for $\PPM\PPM\FF$, and short dashed for $\PPM\PPM$. 
These components are positive (negative) when shown
as thick (thin) curves, and they are abbreviated in Panel~$(c)$ as
PPGG, PPG, PPFG, PPF, and PP. Also note that each component is proportional to
a different combination of the galaxy bias parameters indicated in 
Eqs.~(\ref{eq:powA}), (\ref{eq:powAB}), (\ref{eq:powXX}), (\ref{eq:powAX}),
in addition to the relative velocity bias parameter~$b_r$.
The horizontal gray lines show the diagonal element of the covariance matrix
in the absence of sample variance, approximately equivalent to the shot-noise
contribution (see Sec.~\ref{ssec:cov}). 
It is assumed that the number densities of the tracers are 
$\bng=3\times10^{-4}(\hmpc)^{-3}$ and $\bng=10^{-3}(\hmpc)^{-3}$, 
respectively. 
}
\label{fig:decomp}
\end{figure*}

The auto-power spectrum of the galaxy sample described in Eq.~(\ref{eq:bias})
can be computed \cite{YODASE11} to the second order in power spectrum as
\begin{widetext}
\bear
P_g(\kvec)&=&b_1^2~P_\up{NL}(\kvec)+\int{d^3\qvec\over(2\pi)^3}
P_m(q)P_m(|\kvec-\qvec|)
\bigg[{1\over2}~b_2^2+2~b_1b_2\FF(\qvec,\kvec-\qvec) \nonumber \\
&&+4~b_1b_r\FF(\qvec,\kvec-\qvec)\GG(\qvec,\kvec-\qvec)
+2~b_2b_r\GG(\qvec,\kvec-\qvec)+2~b_r^2\GG(\qvec,\kvec-\qvec)^2\bigg] ~,
\label{eq:powA}
\enar
\end{widetext}
where $P_\up{NL}(\kvec)$ and $P_m(\kvec)$ are the nonlinear and the matter
power spectrum and the relative velocity kernel and the nonlinear matter 
evolution kernel are
\bear
\GG(\kvec_a,\kvec_b)&=&
-{\TT_{u_r}(k_a)\over\TT_m(k_a)}~{\TT_{u_r}(k_b)\over\TT_m(k_b)}~
{\kvec_a\cdot\kvec_b\over k_ak_b}~,\\
\FF(\kvec_a,\kvec_b)&=&{5\over7}+{2\over7}
\left({\kvec_a\cdot\kvec_b\over k_ak_b}\right)^2
+{\kvec_a\cdot\kvec_b\over2}\left({1\over k_a^2}+{1\over k_b^2}\right)~.
\nonumber
\enar
We use the {\footnotesize HALOFIT}
\cite{SMPEET03} to compute the nonlinear matter power
spectrum.
The derivation and the computation of the power spectrum in Eq.~(\ref{eq:powA})
can be found in Appendix~A of \citet{YODASE11}. Apparent from the 
proportionality of the galaxy bias parameters, each term in Eq.~(\ref{eq:powA})
represents the auto- and the cross-contributions
of the nonlinear matter evolution
($\sim b_1$), nonlinear galaxy bias ($\sim b_2$), and the relative velocity
effect ($\sim b_r$).

Figure~\ref{fig:decomp}$a$ illustrates the galaxy power spectrum and the
contributions of its individual components in Eq.~(\ref{eq:powA}).
As our fiducial model, we assume that the galaxy sample has bias parameters
$(b_1,b_2)=(2,1)$ and the relative velocity bias parameter $b_r=0.04$.
The horizontal gray line indicates the approximate level of the shot-noise
contribution $\sqrt{2}/\bng$ to the galaxy power spectrum, where we
assume the galaxy number density is $3\times10^{-4}(\hmpc)^{-3}$. The factor
$\sqrt{2}$ arises, as we plot the power spectrum variance in the absence of 
sample variance (see Sec.~\ref{ssec:cov} for details).
The major contribution to the galaxy power spectrum $P_g(k)$ (thick solid)
is the nonlinear matter power spectrum (thick dotted) on all scales shown in 
Figure~\ref{fig:decomp}. However, the individual components shown as different
curves contribute to the galaxy 
power spectrum at various level at different scales,
given the fiducial values of the bias parameters. Dashed and dotted curves
show the contributions of the nonlinear galaxy bias that are proportional
to $b_2^2$ and $b_1b_2$, respectively. While their contributions can be
as large as 10\% of the total power spectrum at $k\sim0.3\hmpci$, 
they are largely featureless and smooth, which can be readily 
modeled as a smooth broad-band power.

The relative velocity effects are illustrated as solid, dot-dashed and long
dashed curves that are respectively proportional to $b_r^2$, $b_2b_r$, and
$b_1b_r$. The relative velocity power spectrum (solid) is significant on
large scales, amounting to more than 10\% at $k<0.01\hmpci$, but it declines
as $k^{-4}$ on smaller scales, where the BAO peak position is best measured.
This effect is still an important factor at the percent level, determining
the shift in the BAO peak position (see Sec.~\ref{ssec:shift}). However,
for the relative velocity bias larger than the fiducial value $b_r/b_1=0.02$, 
the auto contribution of the relative velocity effect dominates over the
other contributions on all scales. Similarly, the dot-dashed curve represents
the coupling of the relative velocity effect and the nonlinear galaxy bias,
in proportion to $b_2b_r$ and in a similar shape. The oscillation structure
of the both contributions arises due to the relative velocity kernel $\GG$,
and its convolution with the matter power spectrum on large scales
becomes constant as $\GG(\kvec,\kvec)\propto k^{-2}$. 
Last, the long dashed curve
represents the contribution of the relative velocity effect and the 
nonlinear matter evolution, constituting the dominant contribution 
of the relative velocity effect at the
BAO scale. Due to the nonlinear matter kernel $\FF$, this contribution
declines rather slowly with wavenumber, while it retains the oscillation
structure. 

All of the relative velocity effects and the nonlinear effects
are at the percent level or smaller on most scales, given that the total
galaxy
power spectrum (thick solid) is largely determined by the matter power spectrum
(thick dotted). 
However, with high precision measurement of the BAO peak position,
small but non-negligible contributions with oscillation structure can
shift the BAO peak position. In Sec.~\ref{ssec:shift},
we quantify the BAO peak shift in the galaxy power spectrum due to the
relative velocity effect.

\subsection{Cross Power Spectrum}
\label{ssec:cross}
In the past few decades, there have been rapid developments in large-scale
galaxy surveys, and the sheer number has dramatically increased
(e.g., see \cite{WEMOET12}). In this golden age of large-scale surveys,
it is quite common and desirable that two separate and
independent surveys are designed to cover, albeit partially, the same sky
area, generating the synergy effect and providing ways to check systematic
errors.

Here we consider another galaxy sample measured in the survey: Low-mass
star-forming galaxies that form at low redshift after generations of old stars
evolved away long time ago. 
The relative velocity effect in those galaxy samples
is substantially diluted or nearly absent. 
To avoid confusion, we use super- or sub-script~$A$ and~$B$
to refer to each galaxy sample. The galaxy sample~$A$ representing massive
old and red galaxies with the 
relative velocity effect is described by Eq.~(\ref{eq:bias}).
We model the low-mass galaxy sample~$B$ 
with another set of bias parameters
\beeq
\delta_g^B(\xvec)=b_1^B~\delta_m(\xvec)+{b^B_2\over2}~\left[\delta_m^2(\xvec)-
\sigma_m^2\right]~,
\label{eq:bias2}
\eneq
where the relative velocity bias parameter is set $b_r^B=0$.
The auto-power spectrum of the galaxy sample~$B$ can be readily computed
by using Eq.~(\ref{eq:powA}) with the bias parameters replaced by 
($b^B_1$, $b^B_2$). With two galaxy samples in the survey,
we can measure their cross-power spectrum, 
\begin{widetext}
\bear
P_{AB}(\kvec)&=&b^A_1b^B_1~P_\up{NL}(\kvec)
+\int{d^3\qvec\over(2\pi)^3}P_m(q)P_m(|\kvec-\qvec|)
\bigg[{1\over2}b^A_2b^B_2+(b^A_1b^B_2+b^A_2b^B_1)\FF(\qvec,\kvec-\qvec)
\nonumber \\ 
&&  +2b^A_rb^B_1\FF(\qvec,\kvec-\qvec)\GG(\qvec,\kvec-\qvec)
+b^A_rb^B_2~\GG(\qvec,\kvec-\qvec)\bigg] ~.
\label{eq:powAB}
\enar
\end{widetext}

Figure~\ref{fig:decomp}$b$ depicts the cross-power spectrum $P_{AB}(k)$
of the galaxy
sample~$A$ in Fig.~\ref{fig:decomp}$a$ and the galaxy sample~$B$, where
we assume the bias parameters for the galaxy sample~$B$ are
$(b^B_1,b^B_2)=(1,-0.4)$ and the number density is $10^{-3}(\hmpc)^{-3}$.
Low-mass star forming galaxies are known to be less biased but more abundant
(e.g., see \cite{BLJUET09}). 
Since we sample two discrete but distinct objects, there
is no shot-noise contribution to the cross-power spectrum $P_{AB}(k)$.
However, there exist the shot-noise contributions from each galaxy sample
to the power spectrum variance, and the horizontal
gray line shows this contribution $\sqrt{1/\bng^A\bng^B}$ in the absence
of sample variance.

Similarly to the auto-power spectrum $P_{AA}(k)$ in Fig.~\ref{fig:decomp}$a$,
the dominant contribution to the cross-power spectrum (thick solid)
is the matter power spectrum (thick dotted). The contributions of the 
nonlinear galaxy bias ($b_2^Ab_2^B$; dashed) and its coupling with the 
nonlinear matter evolution ($b_1^Ab_2^B+b_2^Ab_1^B$; dotted) 
are also in similar shape, but the former is now negative as $b_2^B<0$
for low mass galaxies. In regard to the relative velocity contributions
to the cross-power spectrum, the notable difference compared to the
auto-power spectrum $P_{AA}(k)$ is the absence of the
relative velocity auto-power spectrum. While the coupling of the
relative velocity effect and the nonlinear galaxy bias $\propto b_r^Ab_2^B$
has a negative sign in the cross power spectrum, the contribution of the
relative velocity effect and the nonlinear matter evolution 
$\propto b_r^Ab_1^B$ (long dashed) remains the dominant factor 
with same sign at the BAO scale as in $P_{AA}(k)$.

Compared to the auto-power spectra $P_{AA}(k)$ and $P_{BB}(k)$, the 
cross-power spectrum $P_{AB}(k)$ in Fig.~\ref{fig:decomp}$b$ provides
a new way to constrain the relative velocity effect with signal-to-noise
ratio as large as their auto power spectra. However, as two galaxy samples
probe the same underlying matter distribution in the survey, 
the auto- and the cross-power
spectra are {\it not} independent, and their covariance matrix needs to
be properly taken into consideration (see Sec.~\ref{ssec:cov}).

\subsection{Multi-Tracer Analysis}
\label{ssec:multi}
Since different galaxy samples
depend on the same underlying matter distribution, 
their stochastic nature shares the common origin. Based on this observation,
the multi-tracer technique is developed
\cite{SELJA09} to remove the intrinsic stochasticity
by taking ratios of multiple tracers. With shot-noise present in
practice, the gain in signal-to-noise ratio is much less than in an
idealized situation, but a significant gain is still achievable (e.g.,
see \cite{HASEDE11,HASEDE12,CABESH11,YOHAET12}), especially when
combined with the shot-noise cancelling technique \cite{SEHADE09,HASEET10}.

Here we construct a particular combination~$X$, out
of the two galaxy samples~$A$ and~$B$ to eliminate the sample variance
\bear
\label{eq:mulng}
n_g^X&\equiv& V_s\left[b_1^A\bng^A n_g^B(\xvec)-b_1^B\bng^B n_g^A(\xvec)\right]
\\
&=&\bng^X\left(1+\delta_X\right)+\varepsilon_X~,\nonumber
\enar
where $\varepsilon_X$ is the shot-noise, 
$\bng^A$ and $\bng^B$ are the mean number densities 
of each galaxy sample and $V_s$ is the survey volume. As each galaxy sample
is composed of its mean and fluctuation, the
combination~$X$ can also be decomposed as the mean number density
\beeq
\bng^X=V_s\bng^A\bng^B\left(b_1^A-b_1^B\right)~,
\eneq
and its fluctuation around the mean
\bear
\label{eq:delX}
\delta_X(\xvec)&=& {b^A_1\delta^B_g-b^B_1\delta^A_g\over b_1^A-b_1^B} \\
&=&{1\over2}~{b^A_1b^B_2 - b^B_1b^A_2\over b_1^A-b_1^B}
~\delta_m^2(\xvec)-{b^B_1b^A_r\over b_1^A-b_1^B}~u_r^2(\xvec)~.\nonumber
\enar
The sample variance is explicitly
eliminated in Eq.~(\ref{eq:delX}) to the leading order,
though much smaller stochasticity, next to the leading order terms, remains.
The power spectrum of the sample-variance cancelling combination is then
\begin{widetext}
\beeq
P_{XX}(\kvec)
=\int{d^3\qvec\over(2\pi)^3}P_m(q)P_m(|\kvec-\qvec|)
\bigg[{1\over2}{\left(b^A_1b^B_2-b^A_2b^B_1\right)^2\over(b_1^A-b_1^B)^2}
+{2\left(b^A_1b^B_2-b^B_1b^A_2\right)b^A_rb^B_1\over(b_1^A-b_1^B)^2}
~\GG(\qvec,\kvec-\qvec)
+{2(b^A_rb^B_1)^2\over(b_1^A-b_1^B)^2}~\GG^2(\qvec,\kvec-\qvec)\bigg] ~.
\label{eq:powXX}
\eneq
\end{widetext}
The sample-variance cancelling combination has different shot-noise 
contribution, as it is the combination of two distinct galaxy samples.
Assuming the Poisson shot-noise for each sample,
the shot-noise contribution to the power spectrum $P_{XX}(k)$ is 
\beeq
N_X\equiv{\AVE{\varepsilon_X\varepsilon_X}\over(\bng^X)^2}
={(b_1^A)^2\bng^A+(b_1^B)^2\bng^B\over\bng^A\bng^B(b_1^A-b_1^B)^2}~.
\eneq

Figure~\ref{fig:decomp}$c$ illustrates the 
auto-power spectrum of the sample-variance cancelling combination.
Since the leading contribution of the matter power spectrum is eliminated, 
there exist no contributions of the coupling with the nonlinear matter
evolution, and 
the amplitude of the power spectrum (thick solid) is overall smaller than 
those
shown in the upper panels. The contribution of the nonlinear galaxy bias
(dashed) is the dominant contribution on all scales, except on large scales
where the relative velocity contributions (solid and dot-dashed) become
more important. The advantage of using the sample-variance
cancelling combination is that the ratio of the relative velocity effect 
to the total power spectrum is greatly enhanced and the oscillation structure
around $k\sim0.1\hmpci$ is largely due to the relative velocity effect,
as the leading-order matter density contribution is eliminated.
However, the gray horizontal lines show the power spectrum variance
 $\sqrt{2}N_X$ in case $P_{XX}=0$, and the shot-noise contribution to the
power spectrum $P_{XX}(k)$ is non-negligible except on large scales.

With the combination~$X$ in addition to two galaxy samples~$A$ and~$B$, we can
also construct the cross-power spectrum of the sample-variance cancelling
combination and the galaxy sample
\begin{widetext}
\bear
\label{eq:powAX}
P_{AX}(\kvec)&=&\int{d^3\qvec\over(2\pi)^3}P_m(q)P_m(|\kvec-\qvec|)
\bigg[{1\over2}{b^A_2(b^A_1b^B_2-b^B_1b^A_2)\over b_1^A-b_1^B}
+{b^A_1(b^A_1b^B_2-b^B_1b^A_2)\over b_1^A-b_1^B}\FF(\qvec,\kvec-\qvec)  \\
&&-{2b^A_1b^B_1b^A_r\over b_1^A-b_1^B}
\FF(\qvec,\kvec-\qvec)\GG(\qvec,\kvec-\qvec)
+{b^A_r\left(b^A_1b^B_2-2b^B_1b^A_2\right)\over b_1^A-b_1^B}
~\GG(\qvec,\kvec-\qvec)
-{2b^B_1(b^A_r)^2\over b_1^A-b_1^B}~\GG^2(\qvec,\kvec-\qvec)\bigg] ~,
\nonumber
\enar
\end{widetext}
and the shot-noise contribution to the cross-power spectrum $P_{AX}(k)$ is 
\beeq
N_{AX}\equiv{\AVE{\varepsilon_A\varepsilon_X}\over\bng^A\bng^X}
={b_1^B\over\bng^A(b_1^B-b_1^A)}~.
\eneq
Figure~\ref{fig:decomp}$d$ shows the cross-power spectrum $P_{AX}(k)$
of the sample-variance cancelling combination with the galaxy sample~$A$. 
The cross-power spectrum $P_{AX}(k)$ is similar to the auto-power
spectrum $P_{XX}(k)$, as it is constructed from the sample-variance cancelling 
combination, while it retains some of the cross-contributions
from the galaxy sample~$A$.
As the combination $n_g^X(\xvec)$ in Eq.~(\ref{eq:mulng}) is constructed with
negative sign of the galaxy number density $n_g^A(\xvec)$,
all the coefficients of the bias parameter combination in Eq.~(\ref{eq:powAX})
are negative, and we plot $-P_{AX}(\kvec)$. The relative velocity effect 
is again prominent on large scales, where the signal is somewhat larger than
the shot-noise contribution
$\sqrt{N^2_{AX}+N_X/\bng^A}$, shown as the gray horizontal line.

\begin{figure*}
\centerline{\psfig{file=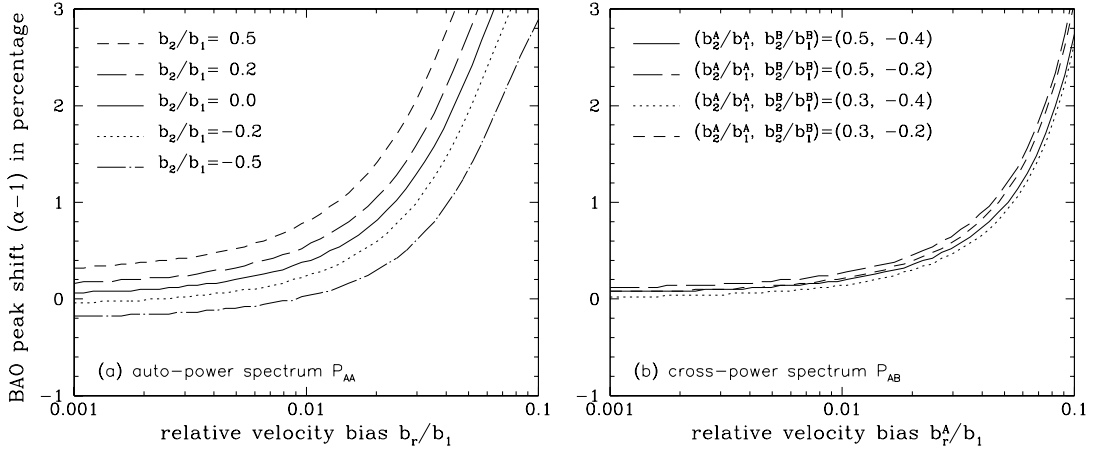, width=3.in, angle=-90}}
\caption{BAO peak shift in the auto-power spectrum (left) and the
cross-power spectrum (right) due to the relative velocity effect. As a 
function of relative velocity bias parameter $b_r/b_1$, the peak shift is
obtained by fitting the template power spectrum in Eq.~(\ref{eq:temp}) to the
full power spectrum, following the  procedure in \cite{SESIET08}.
{\it Left:}~Peak shifts in the auto-power spectrum $P_{AA}(k)$
computed by using Eq.~(\ref{eq:powA}) for various values of
the nonlinear galaxy
bias parameter $b_2/b_1$, given the relative velocity effect
$b_r/b_1$. {\it Right:}~Peak shifts in the cross-power spectrum $P_{AB}(k)$
 computed by using Eq.~(\ref{eq:powAB}) 
for various combination of the nonlinear bias
parameters $b_2/b_1$ for two different galaxy samples~$A$ and~$B$. 
The relative velocity effect is assumed to be present only
in the galaxy sample~$A$, but absent in the galaxy sample~$B$.
}
\label{fig:shift}
\end{figure*}

\subsection{Covariance Matrix}
\label{ssec:cov}
Out of two galaxy samples in the survey, we can construct three power spectra
$P_{AA}$(k), $P_{BB}(k)$, and $P_{AB}(k)$. Since all of them
 are tracing the same underlying
matter distribution, three power spectra are correlated, and its correlation
is described by the covariance matrix. The sample variance cancelling 
combination in Eq.~(\ref{eq:delX})
can be obtained by diagonalizing the covariance matrix of two fluctuations
$\delta_g^A$ and $\delta_g^B$.
To compute the covariance matrix, we first define our estimators for the
power spectra of the two galaxy populations:
\beeq
\hat P_{AA}(\kvec)={1\over V_s}~\delta^A_g(\kvec)\delta^{A*}_g(\kvec)-
{1\over\bng^A}~,
\label{eq:powest}
\eneq
where $V_s$ is the survey volume and 
$\bng^A$ is the number density of the galaxy sample~$A$. Assuming the
Poisson shot-noise, the power spectrum estimator in Eq.~(\ref{eq:powest})
is unbiased $\AVE{\hat P_{AA}(\kvec)}=P_{AA}(\kvec)$,
and a similar estimator can be defined for the galaxy sample~$B$.
In addition, an estimator for the cross power spectrum is 
\beeq
\hat P_{AB}(\kvec)={1\over V_s}~\delta^A_g(\kvec)\delta^{B*}_g(\kvec)~,
\label{eq:croest}
\eneq
and there is no shot-noise contribution to the cross power spectrum.

Given the power spectrum estimators in Eqs.~(\ref{eq:powest}) 
and~(\ref{eq:croest}), the covariance matrix can be readily computed as
\beeq
\up{Cov}\left[\hat P_i(\kvec)\hat P_j(\kvec')\right]
=\left\langle\hat P_i(\kvec)\hat P_j(\kvec')\right\rangle-
P_i(\kvec)P_j(\kvec')~,
\eneq
where $i,j=AA$, $BB$, $AB$. With many independent Fourier modes added
to estimate the power spectra, the power spectrum estimators closely
follow a Gaussian distribution, with which we can analytically derive the
covariance matrix. The diagonal components of the covariance matrix are
\bear
\label{eq:covdia}
\up{Cov}\left[\hat P_{AA}\hat P_{AA}\right]
&=&2\left(P_{AA}+{1\over\bng^A}\right)^2~,\\
\up{Cov}\left[\hat P_{BB}\hat P_{BB}\right]
&=&2\left(P_{BB}+{1\over\bng^B}\right)^2~,\nonumber \\
\up{Cov}\left[\hat P_{AB}\hat P_{AB}\right]
&=&P_{AB}^2+\left(P_{AA}+{1\over\bng^A}\right)
\left(P_{BB}+{1\over\bng^B}\right)~,\nonumber
\enar
where we suppressed the $\kvec$-dependence as each Fourier mode is 
largely independent
and the covariance matrix is nearly diagonal in Fourier space.
The off-diagonal components of the covariance matrix are
\bear
\label{eq:covoff}
\up{Cov}\left[\hat P_{AA}\hat P_{BB}\right]
&=&2\left(P_{AB}\right)^2~,\\
\up{Cov}\left[\hat P_{AA}\hat P_{AB}\right]
&=&2P_{AB}\left(P_{AA}+{1\over\bng^A}\right)~,\nonumber \\
\up{Cov}\left[\hat P_{BB}\hat P_{AB}\right]
&=&2P_{AB}\left(P_{BB}+{1\over\bng^B}\right)~.\nonumber 
\enar
For the sample-variance cancelling combination~$X$, we have two estimators
\bear
\hat P_{XX}(\kvec)&=&{1\over V_s}~\delta_X(\kvec)\delta^{*}_X(\kvec)-N_X~,\\
\hat P_{AX}(\kvec)&=&{1\over V_s}~\delta^A_g(\kvec)\delta^{*}_X(\kvec)-
N_{AX}~,
\enar
and their covariance matrix 
\bear
\up{Cov}\left[\hat P_{XX}\hat P_{XX}\right]
&=&2\left(P_{XX}+N_X\right)^2~, \\
\up{Cov}\left[\hat P_{AX}\hat P_{AX}\right]
&=&\left(P_{AX}+N_{AX}\right)^2\\
&&+\left(P_{AA}+{1\over\bng^A}\right)\left(P_{XX}+N_X\right)~.\nonumber
\enar

While the stochastic nature yields irreducible variance of the estimators
 described by the
covariance matrix, their mean can be more accurately estimated by measuring
independent Fourier modes available in the survey. 
The number of Fourier modes in a given survey is often estimated
\cite{FEKAPE94} as
\beeq
N_k={4\pi k^2\Delta k\over(2\pi)^3}\int_{V_s}d^3\xvec~
\left[{\bng P_g(\kvec)\over1+\bng P_g(\kvec)}\right]^2\simeq
{4\pi k^2\Delta k\over(2\pi)^3}~V_s~,
\eneq
where $\Delta k$ is the bin width in Fourier modes and we assumed the
galaxy samples are sample-variance limited. Since the covariance matrix
is computed accounting for the wavevectors $\kvec$ and $-\kvec$, the number
$N_k$
of wavemodes appears larger by a factor~2 than in the usual calculation.

\subsection{Shift in the BAO Peak Position}
\label{ssec:shift}
Here we quantify the shift in the BAO peak position from the galaxy power
spectrum measurements in the presence of the relative velocity effect.
Due to the nonlinear effects, the BAO peak is broadened and shifted at the
sub-percent level \cite{SESIET08,SHZA12}. However, these scale-dependent
growth and anomalous nonlinear power can be modeled and marginalized over
in measuring the BAO peak
position. Following \citet{SESIET08}, we adopt the template power spectrum
to account for the scale-dependent growth and the broad-band power
\beeq
\label{eq:temp}
P_t(k)=\left(\sum_{i=0}^2c_ik^i\right)P_\up{evo}\left({k\over\alpha}\right)
+\sum_{i=0}^7a_ik^i~,
\eneq
where three coefficients~$c_i$ and eight coefficients~$a_i$ are taken as
free parameters. Accounting for the damping of the BAO peaks in the linear
matter power spectrum , the evolved matter power spectrum is computed as
\beeq
P_\up{evo}(k)=[P_\up{lin}(k)-P_\up{nw}(k)]
\exp\left(-{k^2\Sigma_m^2\over2}\right)+P_\up{nw}(k)~,
\label{eq:evo}
\eneq
where $P_\up{nw}(k)$ is the fit to the broad-band power of the linear
matter power spectrum \cite{EIHU98} and $\Sigma_m=6.6~\hmpc$ is
the degradation parameter of the BAO wiggles at $z=0.57$
\cite{SESIET08}. The template
power spectrum $P_t(k)$ is fitted to the galaxy power spectrum with the
relative velocity effect, and any deviation of~$\alpha$ from unity indicates
the shift in the BAO peak position.

Figure~\ref{fig:shift} shows the shift parameters~$\alpha$ 
in the BAO peak position
of the galaxy power spectrum (left) as a function of galaxy
bias parameters and its
cross-power spectrum (right)
with another galaxy sample without the relative velocity
effect. The best-fit shift parameter~$\alpha$ is obtained by fitting the
template power spectrum in Eq.~(\ref{eq:temp}) to the galaxy power spectrum
in Eq.~(\ref{eq:powA}) or the cross-power spectrum in Eq.~(\ref{eq:powAB})
over a range $0.02\hmpci<k<0.35\hmpci$. 
In the absence of the relative velocity effect, the shift in the BAO
peak position is at the sub-percent level for various values of
the nonlinear galaxy bias parameter~$b_2$. 

\begin{figure*}
\centerline{\psfig{file=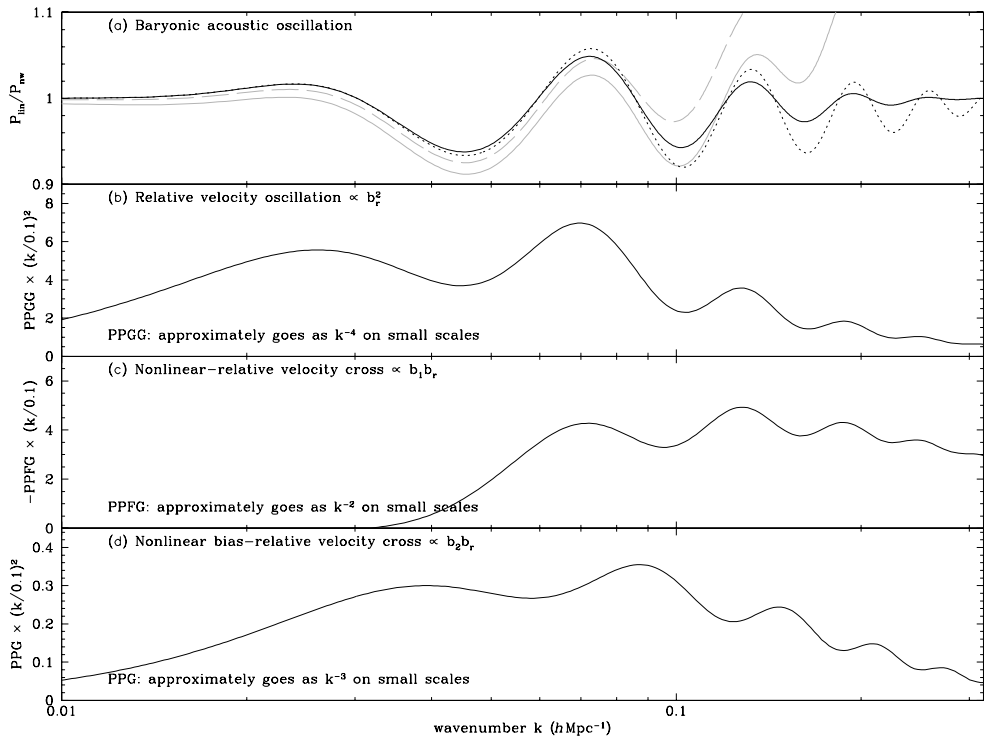, width=4.0in, angle=-90}}
\caption{Baryonic acoustic oscillation and relative velocity contributions
at $z=0$. $(a)$~Ratios to the no-wiggle power spectrum. 
The BAO template in Eq.~(\ref{eq:evo}) is shown as solid curve.
Various curves represent
the linear matter power spectrum (dotted), one-loop SPT power spectrum
(gray dashed), and the {\footnotesize HALOFIT}
nonlinear matter power spectrum (gray solid).
$(b)$, $(c)$, $(d)$~Various 
contributions of the relative velocity effect to the galaxy
power spectrum. The amplitude of each component is normalized 
by the no-wiggle power spectrum
at $k=0.1~\hmpci$ and scaled by various powers of wavenumber to best contrast
their oscillation structure. All the bias parameters are set unity for
comparison.  The contributions PPGG in panel~$(b)$ and PPFG
in panel~$(c)$ are largely in phase with the BAO, while
the contribution PPG is out-of-phase.
}
\label{fig:oscillation}
\end{figure*}

For various values of the nonlinear galaxy bias parameters, the shift
in the BAO peak position due to the relative velocity effect is less
than a percent at $b_r/b_1\leq0.01$ for both the auto-power and the
cross-power spectra. At a fixed value of
the relative velocity bias parameter, 
the overall level of the BAO peak shift is smaller in the cross-power spectrum
(right) than in the auto-power spectrum (left), simply because 
the galaxy bias parameters for the sample~$B$ are smaller than for 
the sample~$A$ and there is no relative velocity effect in the second
tracer. Furthermore, the general trend of the BAO peak shift as a function of
the relative velocity bias parameter is similar for both cases, since
the dominant contribution to the BAO peak shift over the fitting range
arises from the coupling of the nonlinear matter evolution and the relative
velocity effect shown as long dashed curves in Fig.~\ref{fig:decomp} and
their sign remains unchanged in the cross-power spectrum, shifting the
BAO peak position to the same direction as in the auto-power spectrum.

For large values of the relative velocity bias parameter,
a substantial shift in the BAO peak position may occur, but the auto-power
spectrum is better suited for measuring the peak shift due to the relative
velocity effect than the cross-power
spectrum, when the other galaxy sample
is expected to have no relative velocity effect.
However, with two galaxy samples, 
we can construct the sample-variance cancelling
combination in Eq.~(\ref{eq:delX}) and measure its power spectrum
in Fig.~\ref{fig:decomp}$c$ to visually identify the relative velocity effect.
The fitting procedure of \citet{SESIET08} is designed to remove the
broad-band shape and isolate the BAO wiggles in the usual galaxy power 
spectrum, and hence it cannot be used to quantify the shift in the BAO
peak position of the power spectrum of the sample-variance cancelling
combination in Fig.~\ref{fig:decomp}$c$ or its cross-power spectrum in
Fig.~\ref{fig:decomp}$d$.

Figure~\ref{fig:oscillation} further illustrates the oscillation structure
of various relative velocity contributions to the galaxy power spectrum.
Compared to the linear matter power spectrum (dotted) in 
Fig.~\ref{fig:oscillation}$a$, gray curves show the one-loop
SPT power spectrum (dashed) and the {\footnotesize HALOFIT}
nonlinear matter power spectrum
(solid). Though they appear to have quite different oscillation structure,
these nonlinear matter power spectra generally yield negligible
shift in the BAO peak position
 as in Fig.~\ref{fig:shift}, when they are fitted with the
the evolved matter power spectrum $P_\up{evo}(k)$ (solid), 
after marginalizing
over the nonlinear scale-dependent growth and the anomalous power
with free parameters in Eq.~(\ref{eq:temp}).

Figures~\ref{fig:oscillation}$b-$\ref{fig:oscillation}$d$ compare the
relative velocity contributions that are fitted with $P_\up{evo}(k)$
in Fig.~\ref{fig:oscillation}$a$. To remove the broad-band power of the
relative velocity contributions and isolate their oscillation structure,
we apply various powers of wavenumber $k$ normalized at $0.1~\hmpci$ and
divide the contributions to the no-wiggle power spectrum $P_\up{nw}(k)$
at $k=0.1~\hmpci$.
A unity at $k=0.1~\hmpci$ would therefore indicate that its contribution
is as large as the no-wiggle power spectrum, if the corresponding
bias parameters are
set unity. The oscillation structure in the auto-contribution of the relative
velocity effect in Fig.~\ref{fig:oscillation}$b$ is largely in phase
with $P_\up{evo}(k)$, but with $k^{-4}$ scaling and $b_r^2$, 
its contribution is rather limited to large scales $k<0.1~\hmpci$.

Figure~\ref{fig:oscillation}$c$ describes the cross contribution of the
nonlinear matter evolution and the relative velocity effect. With the
weak scaling relation $k^{-2}$ and the amplitude at the percent level 
at $k=0.1\hmpci$ for $b_1\sim1$ and $b_r\sim0.01$, the cross contribution
affects the shift in the BAO peak position most significantly among various
relative velocity contributions over a range $k>0.1~\hmpci$.
The other cross-contribution of the nonlinear galaxy bias and the
relative velocity effect is shown in Fig.~\ref{fig:oscillation}$d$. With the
steep scaling $k^{-3}$ and rather small amplitude, its impact on the
BAO peak shift is rather weak, and the nonlinear galaxy bias parameter
$b_2$ often vanishes in certain galaxy samples.

Compared to our previous calculation of the BAO peak shift in the galaxy 
power spectrum in \citet{YODASE11},\footnote{The nonlinear matter power 
spectrum $P_\up{NL}(k)$ in Eq.~(4.2) of \citet{YODASE11} is
computed by using the one-loop standard perturbation theory (SPT)
$P_\up{NL}(k)=P_m(k)+P_{13}(k)+P_{22}(k)$. 
However, there was a factor~2 error in computing the one-loop correction 
$P_{22}(k)$. With proper calculations of $P_\up{NL}(k)$, 
we find that the shift in the BAO peak position shows the similar trend
as in Fig.~\ref{fig:shift} --- the change of direction in the peak shift for
$b_2<0$ cases in Fig.~5 of \citet{YODASE11} disappeared.
No further significant change arises due to this correction, and
Fig.~4 and~6 of \citet{YODASE11} are largely unaffected.}
we find that the overall trend of the
BAO peak shift is similar. However, since we use
the {\footnotesize HALOFIT}
 \cite{SMPEET03} that better fits the nonlinear matter power
spectrum in simulations than the one-loop SPT power spectrum,
the overall $\chi^2$-value of the fit has dramatically improved over the
previous calculations. As is evident in Fig.~\ref{fig:oscillation}$a$,
the one-loop SPT power spectrum significantly deviates from simulations
at $k>0.2~\hmpci$. Though the evolved matter power spectrum $P_\up{evo}(k)$
has little oscillation at $k>0.2~\hmpci$ shown in 
Fig.~\ref{fig:oscillation}$a$,
we believe that over the fitting range $k=0.02-0.35~\hmpci$,
a more realistic model of the nonlinear matter power spectrum
would be better suited in quantifying the shift in the BAO peak position.

\subsection{Template Power Spectrum
for Relative Velocity Effect}
\label{ssec:temp}

In the presence of the relative velocity effect, 
the BAO peak position in the galaxy power spectrum 
can be shifted as large as a few percents shown in Fig.~\ref{fig:shift},
depending on the relative velocity bias parameter. Since the shift in the BAO
peak position is {\it not} due to the nonlinear scale-dependent growth
around the BAO peak as shown in Figs.~\ref{fig:decomp} 
and~\ref{fig:oscillation}, the evolved linear matter power spectrum 
$P_\up{evo}(k)$ in 
Eq.~(\ref{eq:evo}) is {\it inadequate} in describing the galaxy power spectrum
with the relative velocity effect --- the deviation of~$\alpha$ from unity
in Fig.~\ref{fig:shift} is the evidence of missing physics, {\it the relative
velocity effect}, not the indication of incorrect cosmological parameters
in the fiducial model.

In order to properly account for the relative velocity effect in the galaxy
power spectrum and the correlation function, we construct a different template
power spectrum that can be used to isolate the shift~$\alpha$
in the BAO peak position,
only due to the change in cosmology, not due to the relative velocity effect.
First, we construct a smooth power spectrum $P_s(k)$
that best describes the broad-band
power of the galaxy power spectrum with the relative velocity effect,
but without the oscillation structure, in analogy to the no-wiggle power
spectrum $P_\up{nw}(k)$. Given the galaxy bias parameters
$\bdi{b}=(b_1,b_2,b_r)$, we fit the template power spectrum in 
Eq.~(\ref{eq:temp}) to the galaxy power spectrum $P_g(k|\bdi{b})$
in Eq.~(\ref{eq:powA}) to obtain the peak shift~$\alpha$ and the nuisance
parameters $c_i$ and $a_i$ in Eq.~(\ref{eq:temp}). Second, we construct
the smooth power spectrum as
\beeq
P_s(k|\bdi{b})=\left(\sum_{i=0}^2c_ik^i\right)
P_\up{nw}\left({k\over\alpha}\right)+\sum_{i=0}^7a_ik^i~,
\label{eq:smooth}
\eneq
where the fitted parameters $(\alpha,c_i,a_i)$ depend on the galaxy
power spectrum $P_g(k|\bdi{b})$. The evolved matter power spectrum for
the relative velocity effect is then defined as
\beeq
P_\up{evo}(k|\bdi{b})=\bigg[P_g(k|\bdi{b})-P_s(k|\bdi{b})\bigg]
\exp\left(-{k^2\Sigma_m^2\over2}\right)+P_s(k|\bdi{b})~.
\label{eq:newevo}
\eneq

We repeat the exercise of fitting the template power spectrum $P_t(k)$
in Eq.~(\ref{eq:temp}) to the galaxy power spectrum with the relative velocity
effect as in Fig.~\ref{fig:shift}, 
but in this case, with new evolved power spectrum 
$P_\up{evo}(k|\bdi{b})$ in Eq.~(\ref{eq:newevo}) for each set of galaxy bias
and the relative velocity bias parameters. We find {\it no} meaningful
shift ($\leq0.05\%$)
in all cases, validating the use of the template power spectrum for
the relative velocity effect. 
Further refinement of the evolved power spectrum
$P_\up{evo}(k|\bdi{b})$ can be made by iteratively computing the smooth
power spectrum $P_s(k|\bdi{b})$, until no shift is achieved. However,
we find that the nuisance parameters of the template power spectrum
in Eq.~(\ref{eq:temp}) provide a good description of the broad-band power
of the galaxy power spectrum with the relative velocity effect and
no iteration in most cases is necessary; the maximum shift at one iteration
is $0.4\%$ at $b_r/b_1=0.1$.

\begin{figure*}
\centerline{\psfig{file=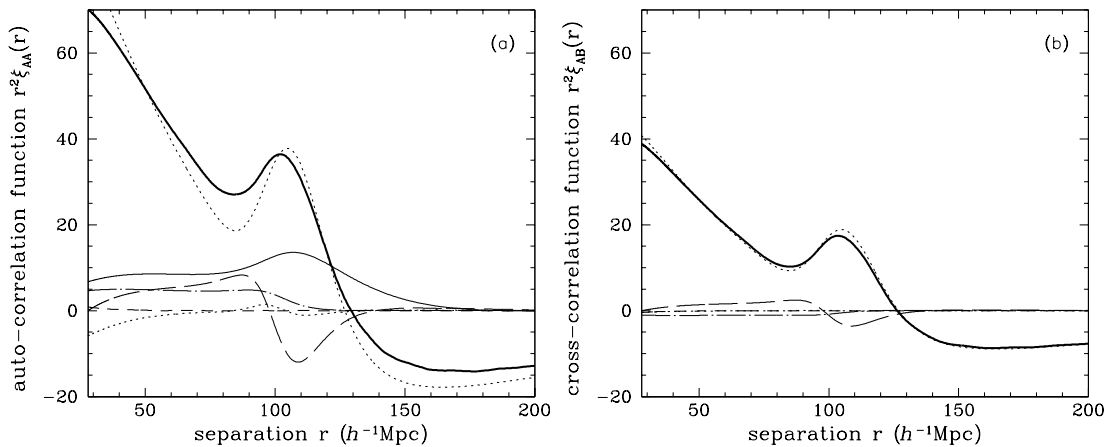, width=3.in, angle=-90}}
\caption{Auto and cross correlation functions of two galaxy samples.
The galaxy sample~$A$ is assumed to have bias parameters
$(b^A_1,b^A_2)=(2,1)$ with the relative velocity effect $b_r^A=0.04$,
and the galaxy sample~$B$ has $(b^B_1,b^B_2)=(1,-0.4)$ without the
relative velocity effect $b_r^B=0$. As a reference,
the dotted curves show the
template correlation function in Eq.~(\ref{eq:correvo}) 
multiplied by $(b^A_1)^2$ and 
$b_1^Ab_1^B$ in each panel, respectively. The best-fit smooth power
spectrum in Eq.~(\ref{eq:smooth}) 
is used to damp the BAO wiggles and compute the full
correlation function (solid). The correlation functions are scaled to match
the template correlation function (dotted) at $r=50\hmpc$.
Individual contributions to the
total correlation functions are shown with line types identical to those
in Fig.~\ref{fig:decomp}, but its amplitude is multiplied by a factor~2
for better illustration. The BAO peak shift due to the relative velocity effect
is more prominent in the auto-correlation function than in the 
cross-correlation function.
}
\label{fig:corr}
\end{figure*}

\subsection{Correlation Function}
\label{ssec:corr}
In addition to the galaxy power spectrum measurements, there exist other
ways to measure the BAO peak position and hence the distance to the galaxy
sample; The galaxy correlation function measurements have been used to
identify the BAO peak position, and the first detection of the BAO scale
in galaxy clustering was made \cite{EIBLET05} by using the correlation 
function measurements. Here we present the relative velocity effect on
the correlation function measurements.

Following \citet{XUPAET12}, we compute the template correlation function,
accounting for the nonlinear growth around the BAO peak 
and the broadening of the BAO feature in the correlation function,
\beeq
\xi_t(r)=B_\xi^2~\xi_\up{evo}(\alpha r)+A_\xi(r)~,
\label{eq:corrtemp}
\eneq
where the constant~$B_\xi$ effectively marginalizes over the correlation
function amplitude, the scale-dependent nonlinear growth around the BAO peak
is fitted with
\beeq
A_\xi(r)={\tilde a_1\over r^2}+{\tilde a_2\over r}+\tilde a_3~,
\label{eq:axi}
\eneq
and the evolved correlation function is the Fourier transform of the
evolved matter power spectrum
\beeq
\xi_\up{evo}(r)=\int{dk~k^2\over2\pi^2}~P_\up{evo}(k)~j_0(kr)
e^{-k^2\sigma_\xi^2}~.
\label{eq:correvo}
\eneq
We adopted the specific functional form in Eq.~(\ref{eq:axi}) to model
the scale-dependent growth around the BAO peak, as in \citet{XUPAET12}.
The cut-off scale in the correlation function is set $\sigma_\xi=1\hmpc$ 
to ensure that the evolved correlation function over the fitting range
$28<r<200~\hmpc$ \cite{ANAUET12} is unaffected by
inadequate modeling of the power spectrum at $k\gg\sigma_\xi$.

For predicting the relative velocity effect in the correlation function,
we compute Eq.~(\ref{eq:correvo}) but with $P_\up{evo}(k)$ replaced by
$P_\up{evo}(k|\bdi{b})$ for the relative velocity effect. However, the best-fit
smooth power spectrum $P_s(k|\bdi{b})$ in Eq.~(\ref{eq:smooth}) is obtained
over a finite fitting
range $k=0.02-0.35~\hmpci$, such that when extrapolated beyond
the fitting range, it fails to provide a good description of the broad-band
shape of the galaxy power spectrum with the relative velocity effect.
Therefore, we use the galaxy power spectrum itself for the smooth power
spectrum outside
the fitting range, noting that the oscillation structure
is weak at $k\leq0.02~\hmpci$ or substantially damped at $k\geq0.35~\hmpci$.

Figure~\ref{fig:corr} shows the correlation function (left) of the
galaxy sample~$A$ and its cross-correlation function (right) with the
galaxy sample~$B$. The dotted curves indicates the 
evolved correlation function in Eq.~(\ref{eq:correvo}), but its amplitude
is multiplied by the galaxy bias parameters $(b_1^A)^2$ and $b_1^Ab_1^B$,
respectively. As shown in Eq.~(\ref{eq:corrtemp}), the evolved correlation
function $\xi_\up{evo}(r)$ is fitted to the correlation function with
nuisance parameters $B_\xi$ and $\tilde a_i$ 
to identify the peak shift~$\alpha$ in the correlation function.
The correlation functions are scaled to match the dotted curves at
$r=50~\hmpc$, effectively setting the nuisance parameter $B_\xi\approx1$.
Compared to the evolved correlation function, the peak position in the
auto-correlation function is shifted to a smaller scale $\alpha>1$, but
its peak is further broadened. These effect are less prominent in the
cross-correlation function. As is shown in Fig.~\ref{fig:shift}, 
the BAO peak shift is $\alpha\sim1\%$ for the auto-power spectrum and
$\alpha\sim0.2\%$ for the cross-power spectrum for the fiducial
galaxy bias parameters and the relative velocity bias parameter.

Various curves in Fig.~\ref{fig:corr} around $\xi=0$ represent the
contributions of the individual components to the correlation function
with line types as in Fig.~\ref{fig:decomp}, but their amplitudes are
multiplied by a factor~2 for illustration. They are computed by Fourier
transforming the each components in Eqs.~(\ref{eq:powA}) and~(\ref{eq:powAB}),
and the damping of the oscillations in these components as in 
Eq.~(\ref{eq:newevo}) is neglected for simplicity. The dominant contribution
to the BAO peak shift is the cross-correlation (dashed)
of the nonlinear matter evolution and the relative velocity effect,
followed by the relative velocity correlation (solid). Since the 
broad-band power
will be removed by the nuisance parameters $\tilde a_i$ in $A_\xi(r)$,
it is the contrast of these components that contributes to the BAO peak shift.

\begin{figure}
\centerline{\psfig{file=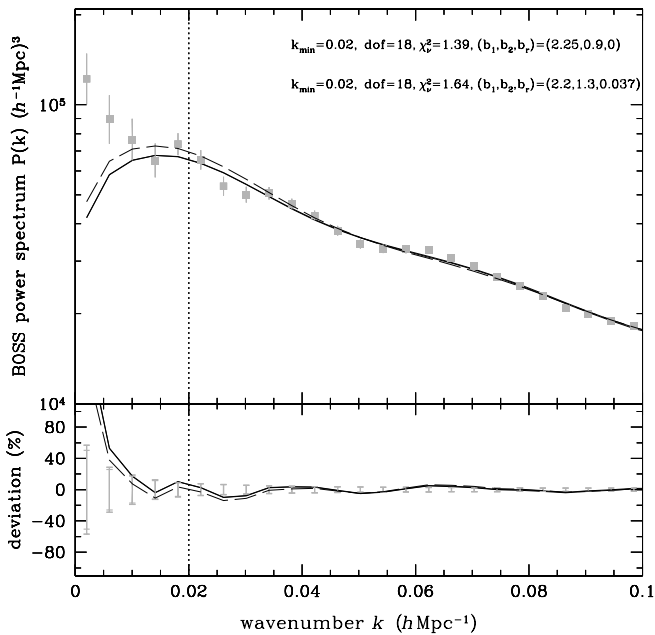, width=3.5in}}
\caption{BOSS power spectrum measurements and the best-fit power spectrum
with the relative velocity effect. The galaxy power spectrum $P_g(k|\bdi{b})$
is fitted to the BOSS measurements (points), accounting for the survey
window function and the integral constraints in Eq.~(\ref{eq:fit})
over the range $k\leq0.1~\hmpci$. Two (nearly identical) solid curves 
show the best-fit galaxy power spectra with the further cut $\kmin=0.02\hmpci$ 
and without $\kmin$. The best-fit model is mainly
determined over the range $k=0.02-0.1~\hmpci$, but the $\chi^2$ difference 
is substantial, depending on whether the measurements at $k<0.02~\hmpci$ 
are included. The dashed curve shows the model with the largest relative
velocity bias parameter that is consistent with the best-fit model
at 1-$\sigma$ level (see Sec.~\ref{sec:boss} for details).  }
\label{fig:boss}
\end{figure}

\section{Constraint on the Relative Velocity Effect from the BOSS}
\label{sec:boss}
Here we use the SDSS-III Data Release~9 (DR9; \citet{AHALET12}) 
galaxy sample, denoted as CMASS sample \cite{WHBLET11}. It comprises
approximately 260,000 galaxies with spectroscopic redshift measurements, 
and their redshift ranges $0.43<z<0.7$ with the effective redshift 
$z_\up{eff}=0.57$ and the number density $\bng=3\times10^{-3}~(\hmpc)^{-3}$. 
The Baryonic Oscillation Spectroscopic Survey (BOSS; \cite{SCWHEI10})
is part of SDSS-III \cite{EIWEET11}, mapping the spatial distribution
of galaxies.
The galaxy power spectrum of the SDSS-III DR9
is measured \cite{ANAUET12} for identifying the BAO peak position and
used for other cosmological purposes (e.g., \cite{ZHSAET12,KASAET13})
and for systematics \cite{ROPEET12}.  We adopt this power spectrum measurements
for our analysis of the relative velocity effect
and refer to it as the BOSS power spectrum measurements.

The BOSS power spectrum is measured by using the traditional 
way to estimate the galaxy power spectrum, known as the FKP 
method \cite{FEKAPE94}. The power spectrum measurements
can be expressed as
the convolution of the underlying galaxy power spectrum with the survey
window function,
\beeq
\hat P_g(\kvec)=\int{d^3\qvec\over(2\pi)^3}P_g(\qvec)
\bigg|\bng^w(\kvec-\qvec)\bigg|^2+P_\up{shot}(k)~,
\label{eq:conv}
\eneq
where $P_\up{shot}(k)$ is the shot-noise contribution and
the survey window function is the Fourier transform of the weighted
mean galaxy number density $\bng^w(\xvec)=w(\xvec)\bng(\xvec)$ and is
normalized to unity
\beeq
1=\int{d^3\kvec\over(2\pi)^3}|\bar n_g^w(\kvec)|^2
=\int d^3\xvec\left[\bar n_g^w(\xvec)\right]^2~.
\eneq
To facilitate the computation 
\cite{COPEET05,PENIET07,PEREET10,BLBRET11,ANAUET12},
the survey window function is often 
angle-averaged and discretized into a window matrix $W[k_i][k_j]$,
such that the three-dimensional integration of the 
convolution in Eq.~(\ref{eq:conv}) becomes a simple matrix multiplication
\beeq
\hat P_g(k_i)=\sum_{k_j}W[k_i][k_j]P_g(k_j)+P_\up{shot}(k_i)~,
\eneq
where the underlying galaxy power spectrum $P_g(k_j)$ is also angle-averaged
and the normalization condition becomes
\beeq
1=\sum_{k_j}W[k_i][k_j]~.
\eneq

Furthermore, since the mean galaxy number density~$\bng$ is unknown and
estimated from the survey itself, the measured fluctuation always vanishes
at the survey scale, even though the underlying fluctuation may not.
This condition, known as the integral constraint, is
implemented by shifting the underlying fluctuation to vanish at the survey
scale \cite{PENI91}. 
Therefore, the measured power spectrum involves additional shift
due to the integral constraint, and our estimate for the galaxy power spectrum
is
\bear
\hat P_g(k_i)&=&\sum_{k_j}W[k_i][k_j]P_g(k_j)+P_\up{shot}(k_i) \nonumber \\
&&-{W(k_i)\over W(0)}\sum_{k_j}W[0][k_j]P_g(k_j)~.
\label{eq:fit}
\enar
Given the galaxy bias parameters $\bdi{b}=(b_1,b_2,b_r)$, we fit the
galaxy power spectrum $P_g(k|\bdi{b})$ in Eq.~(\ref{eq:powA}) to the
BOSS power spectrum measurements, accounting for the survey window function
and the integral constraint in Eq.~(\ref{eq:fit}).
However, for our simple analysis, we keep the cosmological parameters fixed
in fitting the BOSS measurements and constraining the relative velocity
bias parameter~$b_r$.

Figure~\ref{fig:boss} illustrates the best-fit galaxy power spectrum (solid)
with the relative velocity effect from the BOSS measurements (points).
The maximum wavenumber, over which the model is fitted, is set
$\kmax=0.1~\hmpci$.
Our best-fit model prefers {\it no} relative velocity effect in the 
measurements, and the best-fit nonlinear galaxy bias parameters are consistent
with the BOSS measurements \cite{ZHSAET12}, where they find
$(b_1,b_2)=(2.27,1.02)$. While the
best-fit model (solid) provides a good fit to the measurements at 
$k>0.02~\hmpci$, there exist
substantial deviations at $k<0.02~\hmpci$, where the measurements are
significantly higher. Moreover, the chi-square per degree-of-freedom 
$\chi^2_\nu=2.07$ is quite large, demanding further scrutiny.

We repeat the exercise with the minimum wavenumber
$\kmin=0.02~\hmpci$ for the fitting range, removing the
largest scale measurements, where the measurements are most susceptible
to stellar contamination \cite{ROPEET12}. The goodness of the fit dramatically
improves ($\chi^2_\nu=1.39$), while the best-fit model parameters are almost
identical to those obtained without $\kmin$. To further investigate this
systematic error, 
we introduce a nuisance parameter~$s$ for the stellar
contamination and marginalize over the nuisance parameter, following the
procedure in \cite{ZHSAET12}. 
We find no discernible change in $\chi^2_\nu$, compared to that without
the nuisance parameter~$s$. However, the best-fit value of~$s$ is 
a few-$\sigma$ away from $s=0$, indicating the existence of non-negligible
stellar contamination at $k\leq0.02~\hmpci$.
For illustration, the dashed curve shows the model that is
consistent with our best-fit model ($\Delta\chi^2=2.17^2$) at 
1-$\sigma$ level but has the largest relative velocity bias parameter.
While the relative velocity effect can enhance the power spectrum on large
scales $k<0.02~\hmpci$, it is difficult to explain away the power excess
in the largest-scale measurements and maintain the goodness of the fit at
$k=0.02-0.1~\hmpci$.

\begin{figure}
\centerline{\psfig{file=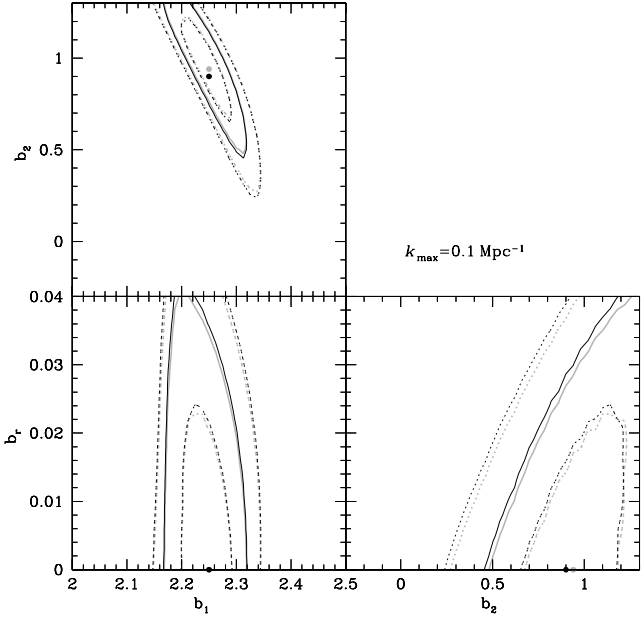, width=3.5in}}
\caption{Constraints on the nonlinear galaxy bias parameters ($b_1$, $b_2$)
and the relative velocity parameter $b_r$ from the BOSS power spectrum
measurements. Contours show 1-$\sigma$ (dotted), 2-$\sigma$ (solid),
3-$\sigma$ (dotted) regions for each two-parameter set. Two nearly overlapping
contours represent the constraints with and without $\kmin=0.02~\hmpci$.
}
\label{fig:cons}
\end{figure}

Figure~\ref{fig:cons} presents the galaxy bias parameter constraints from the
BOSS power spectrum measurements. Contours in Fig.~\ref{fig:cons} depict
$\Delta\chi^2=1.52^2$, $2.49^2$, $3.44^2$ regions, covering 1-$\sigma$,
2-$\sigma$, and 3-$\sigma$ ranges in each parameter space
from the best-fit value indicated by dots. The constraints 
are derived with $\kmin=0.02~\hmpci$, while those shown as
gray contours are with $\kmin=0$. Again, we find no significant difference
in the parameter constraints with different $\kmin$. From the BOSS power
spectrum measurements, we derive the constraint on the relative velocity
bias parameter $b_r<0.033$ at the 95\% confidence level.

The measurements of the BAO peak position provide model-independent
observational constraints on the distance to the galaxy sample (e.g.,
\cite{ANAUET12}). However, the possibility of the relative velocity effect
in the galaxy sample
acts as a systematic error on the BAO measurements, since the BAO peak
position can be shifted if the relative velocity effect is present as
shown in Fig.~\ref{fig:shift}. Given the probability distribution of the
galaxy bias parameters $(b_1,b_2)$ and the relative velocity bias parameter
$b_r$ in Fig.~\ref{fig:cons}, we compute the rms shift in the BAO peak position
\beeq
\langle\Delta\alpha^2\rangle=\int db_1~db_2~db_r~P(b_1,b_2,b_r)~\Delta\alpha^2
(b_1,b_2,b_r)=(0.57\%)^2~,
\eneq
as our estimate of the systematic error in the BOSS measurements of the BAO
peak position, where the shift $\Delta\alpha(b_1,b_2,b_r)$ is shown in 
Fig.~\ref{fig:shift}
and the probability distribution is assumed to be Gaussian 
\beeq
P(b_1,b_2,b_r)\propto\exp\left[-\Delta\chi^2/2\right]~.
\eneq
Given the current constraint on the BAO peak position is 1.6\%
($\alpha=1.033\pm0.017$ \cite{ANAUET12}), the systematic error due to 
the relative velocity effect inflates the uncertainty by 5.5\% 
(hence 1.7\% in total), assuming that the errors are uncorrelated.

\section{Future Measurements of the Relative Velocity Effect}
\label{sec:future}
In Sec.~\ref{sec:boss} we have derived the constraint on the relative
velocity effect from the BOSS measurements. The current constraint is tight,
not only because the measurement precision has dramatically improved,
but also because the cosmological parameters are held fixed.
Looking to the future, we perform a Fisher matrix analysis to
forecast how well future galaxy surveys can constrain the relative velocity
effects and how much improvements can be made by adopting the multi-tracer
analysis, providing guidance to designing future galaxy surveys.

For definiteness, we consider two galaxy samples, 
each of which probe volumes $V=10~(\hgpc)^3$ but with different levels of
overlapping volume. As our fiducial parameters, we adopt that the
galaxy sample~$A$ has the bias parameters $\bdi{b}^A=(b_1,b_2,b_r)=(2,1,0.04)$
and $\bng^A=3\times10^{-4}(\hmpc)^{-3}$, and 
the galaxy sample~$B$ has $\bdi{b}^B=(b_1,b_2)=(1,-0.4)$ and 
$\bng^B=10^{-3}(\hmpc)^{-3}$. We also consider the cosmological parameter
variations parametrized by 
$\bdi{c}=(n_s,\alpha_s,\omega_m,\omega_b,\omega_\up{de},w_0,A_s)$,
where $h^2=\omega_m+\omega_\up{de}$ is the dimensionless
Hubble parameter and we assume
a flat universe with a constant dark-energy equation-of-state.
Therefore, our fiducial model for two galaxy samples is composed of 12
parameters $\pp\equiv(\bdi{c},\bdi{b}^A,\bdi{b}^B)$, where we explicitly
assumed that the relative velocity effect is absent in the galaxy sample~$B$.
Last, as an extension of our model, we introduce two additional free 
parameters for each sample,
accounting for the deviation of the shot-noise power spectrum $P_S(k)$
from the usual Poisson noise $P_S(k)=1/\bng$ 
(e.g., \cite{SEHADE09,HASEET10,BASEET13}). However,
we add a weak prior $\sigma_{P_S}=1/\bng$ on the shot-noise power spectrum
to ensure that the shot-noise
cannot be completely arbitrary.

To estimate the sensitivity of the power spectrum measurements to the
relative velocity effect, we compute the Fisher matrix
\bear
\label{eq:fisher}
F_{\mu\nu}(\pp)&=&\sum_{k=k_\up{min}}^{k_\up{max}}
\sum_{i,j} {1\over N_k ~f^{ij}}\\
&\times&{\partial P_i(k|\pp)\over\partial p_\mu}~
~\up{Cov}\bigg[P_i(k|\pp)P_j(k|\pp)\bigg]^{-1}
{\partial P_j(k|\pp)\over\partial p_\nu}~,\nonumber
\enar
where $\kmin=2\pi/V_s$ and
$i,j=AA$, $BB$, $AB$, representing the auto- and
the cross-power spectra of two galaxy samples~$A$ and~$B$. The fractional 
number of Fourier modes available in the survey is
\beeq
f^{ij}=\bigg\{
\begin{array}{cl} 1 & \quad\up{for}~i=j=AA,~BB\\
V_o/V_s & \quad\up{otherwise}\end{array}\bigg\}~,
\eneq
where $V_o$ is the overlapping volume probed by two galaxy samples.
When two samples probe two independent volumes ($V_o=0$), 
no cross-power spectrum is used in the Fisher matrix calculation.
We also adopt the Planck prior on the cosmological parameter
estimation, following the procedure in \citet{DETF06}. Therefore,
the constraint on the relative velocity bias parameter can be derived as
\beeq
\sigma^2_{b_r}=\left[F^{-1}(\pp)\right]_{b_rb_r}~,
\eneq
after marginalizing over all the cosmological parameters and the remaining
bias parameters.

\begin{figure*}
\centerline{\psfig{file=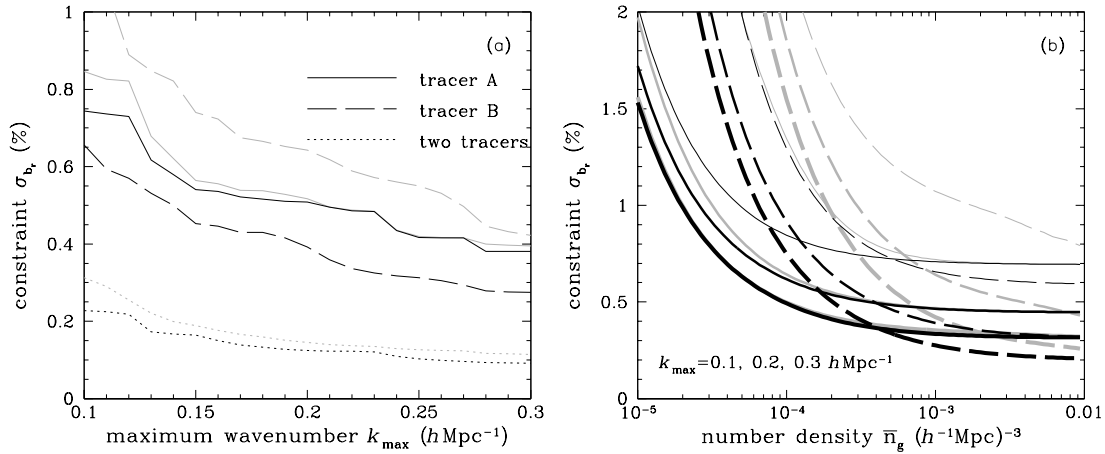, width=3.in, angle=-90}}
\caption{Constraints on the relative velocity parameters in a survey of
volume $V=10~(\hgpc)^3$. The bias parameters of the two galaxy samples are
$(b_1,b_2,b_r)=(2,1,0.04)$ and $(1,-0.4,0)$, and their fiducial values
of number densities are $n_g=3\times10^{-4}$ and $10^{-3}~(\hmpc)^{-3}$,
respectively. {\it Left:} Constraint on the relative velocity bias
 parameter~$b_r$
as a function of $\kmax$. The cosmological parameters and the remaining
bias parameters are marginalized over in deriving the constraint on the
relative velocity effect. Dotted
curve shows the improvement on the constraint by independently measuring 
two galaxy samples 
in the survey (full covariance matrix and their cross-correlation
function are taken into account). Gray curves show the constraints, when we
allow the shot-noise of each sample to be a free parameter.
{\it Right:} Constraint on $b_r$ as a
function of galaxy number density $\bng$. The bias parameters are fixed at
their fiducial values. From top to bottom, curves with different thickness
represent different $\kmax$.}
\label{fig:con}
\end{figure*}

\begin{figure}
\centerline{\psfig{file=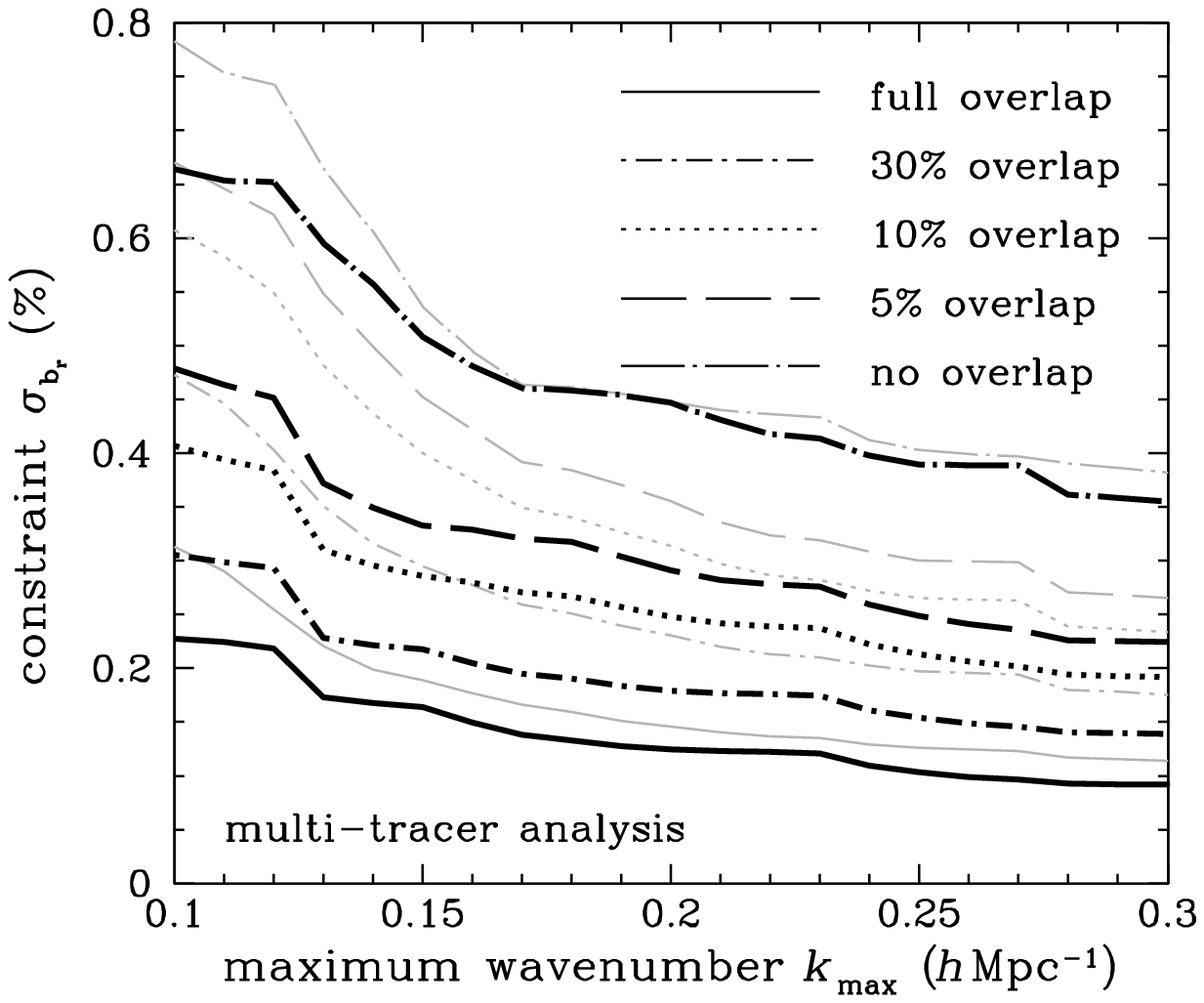, width=3.in, angle=-90}}
\caption{Multi-tracer analysis of the relative velocity effect. The bias
parameters of the two galaxy samples 
are assumed to have the fiducial values as in 
Fig.~\ref{fig:con}. The galaxy sample~$B$ is assumed to be independent of the
relative velocity effect. The survey volume is $V_s=10~(\hgpc)^3$.
The cross-power spectra and the sample-variance cancelling 
combination are only available when the survey volumes overlap,
shown as various
curves. For non-overlapping case, the galaxy samples~$A$ and~$B$ 
probe two independent
survey volumes, and the constraint on the relative velocity effect is derived
from the galaxy sample~$A$ alone, while the galaxy sample~$B$ 
helps constrain the underlying matter distribution. Gray curves represent
the constraints with the same conditions as in black curves,
but we additionally marginalize over the unknown amplitudes of the
shot-noise power spectra.
}
\label{fig:overlap}
\end{figure}

In Fig.~\ref{fig:con}$a$ the solid (galaxy sample~$A$) and the dashed 
(galaxy sample~$B$)
curves show the constraint on the relative velocity bias
parameter from the galaxy power spectrum analysis in a survey of 
volume $V=10~(\hgpc)^{3}$. As we increase the maximum wavenumber $\kmax$,
the constraints are improved in both cases by lifting the degeneracy 
among the cosmological parameters~$\bdi{c}$ and the galaxy bias 
parameters~$\bdi{b}$. The constraint (dashed)
derived from the galaxy sample~$B$ is stronger than the constraint (solid) 
from the galaxy sample~$A$, 
in large because the shot-noise contribution is lower
for the second galaxy sample (with larger number density), but also because
the overall power spectrum amplitude is lower $b_1^B<b_1^A$.
Another 30\% improvements in the constraint arise, if the galaxy bias
parameter~$b_1$
and the matter fluctuation normalization~$A_s$
are known \cite{YODASE11}.

The gray curves show the degradation of the constraints, once we allow the
shot-noise power spectrum to be free. The degradation is minor for the galaxy
sample~$A$, as the relative velocity effect is already present in the galaxy
sample~$A$ and its contributions to the galaxy power spectrum scale quite
differently with a constant shot-noise. However, the dashed curve shows that
there exists a strong degradation for the galaxy sample~$B$.
Since the relative velocity
effect is absent in the galaxy sample~$B$ and the nonlinear galaxy bias 
parameter is negative, the relative velocity contribution can be degenerate
with the shot-noise power spectrum with unknown amplitude
on broad range of scales.

The dotted curve shows the improved constraints, if two galaxy samples
are available
in the same survey area, where not only the auto-power spectrum of each
sample, but also the cross-power spectra in Fig.~\ref{fig:decomp} can
be utilized to constrain the relative velocity bias parameter.
A factor of few improvements on the constraint arise, demonstrating that
the multi-tracer analysis in Sec.~\ref{ssec:multi} extracts more information
than the sum of two independent analysis. A sub-percent level constraint
on the relative velocity bias parameter can be readily achievable, providing
a tight constraint
on the star formation history in the early Universe. With two tracers combined,
the constraint is less sensitive to the uncertainties in the shot-noise
power spectra, especially when the maximum wavenumber is large enough
that the degeneracy among galaxy bias parameters can be lifted.

The right panel illustrates the dependence of the galaxy number density.
We keep the galaxy bias parameters fixed for two galaxy samples, 
and three curves
with different thickness represents different maximum wavenumber $\kmax$.
At low galaxy number density $\bng\RA0$, the galaxy sample~$B$ (dashed) is more
shot-noise limited than the galaxy sample~$A$, as the overall amplitude $b_1^B$
is lower. In the opposite limit $\bng\RA\infty$, 
where the measurements are sample-variance limited, they both provide
equally strong constraint on the relative velocity bias parameter,
when the constraint~$\sigma_{b_r}$ is scaled with $b_1$. The constraint
$\sigma_{b_r}/b_1$ from the galaxy sample~$A$ is slightly better in this limit,
because the galaxy sample~$A$ is assumed to have the relative velocity effect
and the galaxy sample~$B$ is assumed to be devoid of it, 
while the difference is
rather small, as the relative velocity power spectrum $(\sim b_r^2)$
is significant only on large scales.
With unknown amplitude of the shot-noise power spectrum, the constraints
on the relative velocity bias parameter degrade in all cases at various levels.
As the number density increases, most of the information is recovered for
the galaxy sample~$A$, but some degeneracy remains unbroken for the galaxy
sample~$B$.

Figure~\ref{fig:overlap} further investigates the improvement of the
relative velocity constraint by using two independent galaxy samples,
but with
varying degree of overlap in survey volumes. Here we explicitly
assume that the galaxy sample~$B$ is devoid of the relative velocity effect
and the constraint on the relative velocity effect is derived solely by
the galaxy sample~$A$. The dot-dashed curve shows the constraints when there is
no overlapping volume for two galaxy samples. 
In this case, the galaxy sample~$B$
helps constrain the relative velocity effect by providing constraints on
the underlying cosmological parameters and lifting degeneracy in the
constraints derived by the galaxy sample~$A$ (hence the dot-dashed curve should
not be compared to the sum of the solid and the dashed curves in
Fig.~\ref{fig:con}$a$).
As the overlapping volume increases from top to bottom, the constraints on
the relative velocity effect improve, but the most dramatic improvement
arises when the survey volumes start to overlap (dashed curve). 
Once two galaxy samples are available for their cross-power spectra
in an overlapping volume, they can be used to construct the sample-variance
cancelling combination to isolate the relative velocity effect, dramatically
improving the leverage to constrain the relative velocity effect.

Gray curves show the degradation due to the uncertainties in the shot-noise
power spectra. The constraints inflate by approximately 50\% with 
$\kmax=0.1~\hmpci$, while the degradation becomes minor at $\kmax=0.3~\hmpci$,
consistent with the result in Fig.~\ref{fig:con}$a$.
The limitation of the multi-tracer analysis in achieving an ``infinite''
signal-to-noise ratio is the existence of shot-noise. Recently, a new
technique has been developed \cite{SEHADE09,HASEET10} to suppress the
shot-noise contribution by applying mass-dependent weights to the galaxy 
sample. This technique effectively enhances the galaxy number density and
the galaxy bias parameters. For simplicity, we repeat the multi-tracer
analysis with the galaxy number density $n_g^A=10^{-3}~(\hmpc)^{-3}$,
approximately three times larger than our fiducial value and 
the galaxy bias parameters increased by a factor~1.5, as expected from the
shot-noise cancelling technique \cite{HASEET10}, but we kept the
properties of the galaxy sample~$B$ unchanged. With the shot-noise cancelling
technique, the constraint $\sigma_{b^A_r}/b_1^A$
on the relative velocity bias parameter improves, but the improvement is
rather weak, since the number density of the galaxy sample~$A$ is already
large and no substantial gain is achieved by increasing its number density
as in Fig.~\ref{fig:con}$b$.

\section{DISCUSSION}
\label{sec:discussion}
The supersonic relative velocity effect between baryons and dark matter
plays an important role
in the formation of the first stars and the earliest baryonic structure
by suppressing the dark matter halo abundance,
reducing the gas contents in halos, and boosting the minimum halo mass
in which gas can cool and form stars 
\cite{DAPESE10,TSBAHI11,FIBAET12,MCOL12,NAYOGN12}.
Furthermore, these effects could imprint distinct signatures in the 21cm
fluctuation signals and affect the reionization history of the Universe
\cite{DAPESE10}. In particular, the enhancement of the 21cm signals 
at $z\sim20$ due to the relative velocity effect may be observed 
in the existing low-frequency radio arrays \cite{VIBAET12,FIBAET13}.
Despite the large uncertainties inherent in
the nonlinear galaxy formation at low redshifts, it is speculated
\cite{DAPESE10,GRWHET11,YODASE11,FESMDV12}  that
this modulation of the relative velocity effect on the early baryonic
structure might be inherited by some fraction of stars composing
low-redshift massive galaxies or by local patches of intergalactic medium,
in which massive galaxies form at later times.
Parametrizing our ignorance of the relic amplitude of the remaining
relative velocity effect in galaxy samples at low redshifts,
we have studied the impact of the relative velocity between
baryons and dark matter on the large-scale clustering properties of galaxies.

Drawing on the calculation \cite{YODASE11}, we have extended the computation
of the galaxy power spectrum to the multiple galaxy samples and investigated
the synergy effect of the multi-tracer analysis on isolating the relative 
velocity effect in the galaxy samples. With the small amplitude 
($\lesssim1\%$), if any, 
of the relative velocity effect that may persist until today, it is somewhat
difficult to measure the relative velocity effect in the galaxy power spectrum
measurements alone in a model-independent way. However,
the situation changes dramatically, once we have another galaxy sample
to cross-correlate with, especially when the second galaxy sample is known
to have vanishing relative velocity effect, such as in low-mass star forming
galaxies that mainly consist of newly born stars at low redshifts.

Since the acoustic structure of the relative velocity effect in the galaxy
power spectrum is similar but out-of-phase with the acoustic structure
of the matter and the baryon distributions, the relative velocity effect
can shift the BAO peak position by a few percent at a plausible
range of the relative velocity bias parameter $b_r=0\sim0.02$, if the
relative velocity effect is unaccounted for in determining the BAO peak
position. 
With two distinct galaxy samples measured in an overlapping volume of 
the survey, their cross-power spectrum and the 
cross-correlation function can be used to provide an important 
consistency check to the constraints on the relative velocity effect. 
On scales $k=0.02\sim0.2~\hmpci$, 
where the BAO peak position is best measured, the cross-correlation of the
relative velocity effect and the nonlinear matter evolution is the dominant
contribution to the galaxy power spectrum, shifting the BAO peak position
to the same direction in the galaxy auto- and cross-power spectra, although
the shift is often reduced in the cross-power spectrum,
when the second galaxy sample has negative
nonlinear galaxy bias parameter~$b_2$.

More importantly,
the multi-tracer analysis \cite{SELJA09} takes advantage of the fact that
two galaxy samples trace the same underlying matter distribution, and it
allows one to construct a particular combination of two galaxy samples, 
in which the leading contribution of the matter fluctuation is eliminated,
such that the relative velocity effect is the dominant feature in the
power spectrum on
large scales $k<0.1~\hmpci$, providing a model-independent way to verify
the presence of the relative velocity effect in the galaxy sample.
While the leading stochasticity may be eliminated in this sample-variance
cancelling combination, the shot-noise contribution still remains in the
power spectrum measurements.
In order to maximize the utility of the multi-tracer analysis,
it is preferred that the second galaxy sample is abundant in number density
and has large difference $\Delta b_g=b_1^A-b_1^B$ in their bias factors 
(see also \cite{MCSE09}), which enhances the contrast of the relative
velocity effect to the shot-noise contribution in the galaxy power spectrum.

Using the publicly available BOSS power spectrum measurements \cite{ANAUET12}
of the CMASS galaxy sample \cite{WHBLET11} from the SDSS-III Data Release~9
\cite{AHALET12}, we have derived  the constraint on the
relative velocity effect --- the relative velocity bias parameter is 
constrained $b_r<0.033$ at the 95\% confidence level from the BOSS power
spectrum at $k=0.02-0.1~\hmpci$. Though the constraint may be further improved
by extending the fitting range, the goodness-of-the-fit substantially
degrades, as the systematic error in the theoretical modeling 
becomes non-negligible. Hence
we take the conservative estimate of the upper limit
on the relative velocity effect as $b_r<0.033$. The constraint on the relative
velocity bias parameter yields the systematic error 
$\langle\Delta\alpha^2\rangle^{1/2}=0.57\%$
on the BOSS measurements of the BAO peak position at $z=0.57$. 
Compared to the current
observational error $\Delta\alpha=1.7\%$, the systematic error due to the
relative velocity effect is negligible.

Consider a population of the first stars that form at very early time, in 
which the relative velocity effect is the dominant mechanism, modulating
the spatial fluctuation of the stellar population:
$n_\star\simeq\bar n_\star(1+u_r^2)$. The relative velocity bias parameter
is of order unity at early times \cite{DAPESE10}, and $\delta_m\ll1$.
In contrast, typical
galaxy samples including the CMASS sample are composed of ordinary stars
without the relative velocity effect, and those galaxy samples are well
described by the linear bias relation on large scales today:
$n_\odot\simeq\bar n_\odot(1+b~\delta_m)$.\footnote{Despite our notation
($n_\star$, $n_\odot$) for two different populations of 
stars, they represent the galaxy number densitites on large scales that
are entirely made of each stellar population. They should not be confused
with stellar number densities in some local regions.}
Assuming that the CMASS galaxy sample is composed of these two populations
of stars, we can relate the relic amplitude $b_r$ of the
relative velocity effect to the fraction of the first stars in the CMASS galaxy
sample as
\beeq
n^\up{CMASS}_g=
\bar n_\odot(1+b~\delta_m)+\bar n_\star(1+u_r^2)\simeq
\bng(1+b~\delta_m+f_\star u_r^2)~,
\eneq
where we assumed $\bng\simeq\bar n_\odot\gg\bar n_\star$ and
$f_\star=\bar n_\star/\bng\ll1$. The dimensionless relative velocity $u_r$ is
normalized and largely independent of time. 
The constraint on the relative velocity bias
parameter is, therefore, directly linked with the fraction of the first
stars in the CMASS galaxy sample, and we derive the upper
limit on the fraction of the first stars as 3.3\% in the CMASS galaxy sample
at the 95\% confidence level. 

We used
``first stars'' to refer to a stellar population, of which the spatial
distribution is modulated by the relative velocity effect at early times,
as opposed to the ``ordinary stars'' without the relative velocity effect at
late times. Our constraint applies to the fraction of those stars with the
relative velocity effect, not to the fraction of ``the first stars'' 
that are known as the population III stars, as their lifetime is very short
\cite{TESIET97}. In this regard, our toy model is a phenomenological
description, agnostic of the merger history and the stellar evolution.
A more thorough modeling of the formation history would be interesting,
especially when the relative velocity effect is detected, but it is beyond
the scope of the current investigation.

Furthermore,
regarding the interpretation of our upper limit, a few more 
caveats are in order. 
First, we held the cosmological parameters fixed as the best-fit parameter
set derived in the {\it Planck} result \cite{PLANCK13}. 
With the tight constraints on the cosmological parameters,
variations of the BAO peak position are negligible among
permitted sets of cosmological parameters. We suspect that
a full analysis of the galaxy power spectrum with cosmological parameter 
variation is unlikely to change the upper limit. However, we note that
a cosmological parameter set substantially different from our fiducial model
could favor the presence of the relative velocity effect, albeit marginal
($\sigma_{b_r}\sim3\%$), since the BAO scale would be somewhat different 
in the assumed cosmology. Second, the redshift-space
distortion is neglected in the power spectrum analysis. However, on large
scales, the redshift-space distortion only enhances the power by
a factor $1+2\beta/3+\beta^2/5$ in a scale-independent manner, where
$\beta$ is the anisotropy parameter. Since the
relative velocity effect is measured from its scale-dependent oscillations,
modeling of the redshift-space distortion will leave the
derived constraint unchanged, but we suspect that it would reduce our estimate
of the galaxy bias factor by 13\% with $\beta\simeq0.39$ to a more reasonable
value $b_1\simeq2.0$.

Looking to the future, we have forecasted the constraint on the relative 
velocity effect derivable from the multi-tracer analysis in future galaxy
surveys. With two galaxy samples in a survey of $V=10~(\hgpc)^3$, the 
multi-tracer analysis can achieve a 0.1$-$0.2\% level constraint on the
relative velocity bias parameter, improving the constraint 
by a factor of $3\sim5$ compared to the single tracer analysis.
Since future galaxy surveys aim to measure galaxies at higher redshift 
$z\gg0.5$, the impact of the relative velocity effect on the galaxy power
spectrum, if present in the galaxy samples,
could be larger, providing an opportunity to detect 
the relative velocity effect
in galaxy surveys and better understand the galaxy formation history before
reionization.

\acknowledgments
We thank Florian Beutler and Chris Blake for useful comments and 
Will Percival for providing the BOSS power 
spectrum measurements in electronic form. 
This work is supported by the Swiss National Foundation (SNF) under contract
200021-116696/1 and WCU grant R32-10130.
J.Y. is supported by the SNF Ambizione Grant.

Funding for SDSS-III has been provided by the Alfred~P. Sloan Foundation, 
the Participating Institutions, the National Science Foundation, and the 
U.S. Department of Energy Office of Science. The SDSS-III web site is 
{\footnotesize http://www.sdss3.org/}.

SDSS-III is managed by the Astrophysical Research Consortium for the 
Participating Institutions of the SDSS-III Collaboration including the University of Arizona, the Brazilian Participation Group, Brookhaven National Laboratory, University of Cambridge, Carnegie Mellon University, University of Florida, the French Participation Group, the German Participation Group, Harvard University, the Instituto de Astrofisica de Canarias, the Michigan State/Notre Dame/JINA Participation Group, Johns Hopkins University, Lawrence Berkeley National Laboratory, Max Planck Institute for Astrophysics, Max Planck Institute for Extraterrestrial Physics, New Mexico State University, New York University, Ohio State University, Pennsylvania State University, University of Portsmouth, Princeton University, the Spanish Participation Group, University of Tokyo, University of Utah, Vanderbilt University, University of Virginia, University of Washington, and Yale University. 

\vfill

\bibliography{prd.bbl}

\end{document}